\documentclass[11pt]{article}
\usepackage[usenames]{color} 
\usepackage[cmex10]{amsmath}
\usepackage{amssymb}
\usepackage{amsthm}
\usepackage{txfonts}

\newtheorem{cor}{Corollary}
\newtheorem{defn}{Definition}
\newtheorem{lemma}{Lemma}
\newtheorem{thm}{Theorem}
\newtheorem{eg}{Example}
\newtheorem{prop}{Proposition}

\newcounter{cond}

\def\Real{\mathop{\hbox{\mit I\kern-.2em R}}\nolimits}
\def\Zeal{\mathop{\hbox{\mit Z\kern-.29em Z}}\nolimits}
\def\be{\begin{equation}}
\def\ee{\end{equation}}
\def\pf{\noindent {\bf Proof} \ }
\def\endpf{\hfill $\square$}
\def\nn{\nonumber}
\def\l{\label}
\def\r{\ref}
\def\bs{\backslash}
\def\ra{\rightarrow}
\def\ba{\begin{eqnarray*}}
\def\ea{\end{eqnarray*}}

\def\bfx{{\bf x}}

\def\calA{{\cal A}}

\def\calC{{\cal C}}
\def\calD{{\cal D}}

\def\calF{{\cal F}}

\def\calS{{\cal S}}
\def\calT{{\cal T}}

\def\calX{{\cal X}}

\def\tE{\tilde{E}}
\def\tG{\tilde{G}}

\def\tX{\tilde{X}}

\def\"{{\prime\prime}}

\def\bqy{\begin{eqnarray}}
\def\eqy{\end{eqnarray}}
\def\ol{\overline}
\def\Path{{\rm Path}}
\def\Im{{\it Im}}
\def\Im{{\it Im}}
\def\AII{{\cal A}_{{\rm II}}}
\def\TI{{\calT_{\rm I}}}
\def\TII{{\calT_{\rm II}}}
\def\GV'{{G^*(V')}}

\def\bit{\bibitem}

\usepackage{graphicx}

\begin{document}

\newcommand{\dqdef}{\mbox {$  \ \stackrel{def}{=}\    $}  }

\begin{titlepage}
\title{On Information-Theoretic Characterizations of Markov Random Fields and Subfields}
\author{Raymond W. Yeung\thanks{Raymond Yeung and Qi Chen are with 
Institute of Network Coding and Department of Information
Engineering,
The Chinese University of Hong Kong, N.T., Hong Kong.
Email: \{whyeung, qichen\}@ie.cuhk.edu.hk} 
\and 
Ali Al-Bashabsheh\thanks{Ali Al-Bashabsheh is with the Big Data and Brain Computing Center, Beihang University,
No. 37 Xueyuan Road, Haidian District,
Beijing,
Postcode 100191.
Email: entropyali@gmail.com}
\and 
Chao Chen\thanks{Chao Chen is with School of Electronic Engineering, Xidian University, 710071, China.  Email: cchen@xidian.edu.cn}
\and  
Qi Chen\footnotemark[1]
\and 
Pierre Moulin\thanks{Pierre Moulin is with ECE Department and Coordinated Science Laboratory, University of Illinois, Urbana IL 61801.
Email: moulin@ifp.uiuc.edu}
}

\maketitle

\begin{center}
{\bf Abstract} 
\end{center}

\noindent
Let $X_i, i \in V$ form a Markov random field (MRF) represented by an undirected graph $G = (V,E)$, and $V'$ be a subset of $V$.
 We determine the smallest graph that can always represent the subfield $X_i, i \in V'$ as an MRF.  Based on this result, 
we obtain a necessary and sufficient condition for a
subfield of a Markov tree to be also a Markov tree.
When $G$ is a path so that $X_i, i \in V$ form a Markov chain, it is known that 
the $I$-Measure is always nonnegative and the information diagram assumes a very special structure \cite{KawabataY92}.  
We prove that Markov chain is essentially the only  
MRF such that the $I$-Measure is always nonnegative.  
By applying our characterization of the smallest graph representation of a subfield of 
an MRF, we develop a recursive approach for constructing information diagrams for MRFs.
Our work is built on the set-theoretic characterization of 
an MRF in \cite{YeungLY02}.

\end{titlepage}


\section{Introduction}
\l{sec:intro}
A Markov random field (MRF) is often regarded as a generalization of a
one-dimensional discrete-time Markov chain in the sense that the 
time index for the latter is replaced by a space index for the former.
Historically, the study of MRFs stems from statistical
physics.  The classical Ising model, which is defined on a rectangular lattice, 
was used to explain certain empirically observed facts about
ferromagnetic materials.  
In statistics, the dependencies between variables in a contingency table may also be modeled as an MRF \cite{AsmussenE83}. In image processing and computer vision, the dependencies between pixels or image features are also commonly modeled by MRFs \cite{BlakeKR11}. MRFs have also been used in wireless and ad hoc networking \cite{DoyleKDF06, YazirFGGC010, JeonJ10}.
In recent years, MRFs have been used 
as a model for studying social networks \cite{Snijders11,WangKV13} and big data \cite{SandryhailaM14}.

The foundation of the theory of MRFs may be found in
\cite{Preston74} or \cite{Spitzer71} (also see \cite{Kindermann80} and \cite{Lauritzen96}).  
It was described in \cite{Preston74}
that the theory can be generalized to the context of an arbitrary graph.
In this paper, we discuss such MRFs whose random variables are discrete.
Before we present their formulation, we first introduce some notations that are used
throughout the paper.

In this paper, all random variables are discrete.
Let $X$ be a random variable taking values in an alphabet $\calX$.
The probability distribution for $X$ is denoted as 
$\{ p_X(x), x \in \calX \}$, with
$p_X(x) = {\rm Pr}\{X=x\}$.
When there is no ambiguity, $p_X$ is 
abbreviated as $p$.  The {\em support} of $X$, denoted by $\calS_X$,
is the set of all $x \in \calX$ such that $p(x) > 0$.
If $\calS_X = \calX$, we say that $p$ is 
{\em strictly positive}, denoted by $p > 0$.
Otherwise, $p$ contains
zero probability masses, and we say that $p$ is not strictly positive.
Note that probability distributions with zero probability masses are 
in general very delicate, and they need to be handled very carefully
(see Example~\r{haha} below).
All the above notations naturally extend to two or more random variables.

\begin{prop}
For random variables $X, Y$, and $Z$, $X \Perp Z \ | \                       Y$ if 
and only if
\be
p(x,y,z) = a(x,y) b(y,z) 
\ee 
for all $x$, $y$, and $z$ such that $p(y) > 0$, where $a$ is some function of $x$ and $y$
and $b$ is some function of $y$ and $z$.
\l{prop:MC3}
\end{prop}

The example below illustrates the subtlety of conditional independence when
the probability distribution contains zero probability masses.

\begin{eg}
Let $p$ denote the joint distribution of three random variables $X_1, X_2$, and $X_3$.
In this example, we show that 
\be
\left.
\begin{array}{ll}
X_1 \Perp X_2  \ | \ X_3 \\
X_1 \Perp X_3 \ | \ X_2
\end{array}
\right\} \ \Rightarrow \
X_1 \Perp (X_2, X_3) 
\l{avuibva}
\ee
holds if $p > 0$, but does not hold in general.

Assume that $p > 0$.  Then for all $x_1, x_2$, and $x_3$, by $X_1 \Perp X_2 \ | \ X_3$, we have
\be
p(x_1, x_2, x_3) = \frac{p(x_1, x_3) p(x_2, x_3)}{p(x_3)} ,
\l{aiugbvv}
\ee
and by  $X_1 \Perp X_3 \ | \ X_2$, we have
\be
p(x_1, x_2, x_3) = \frac{p(x_1, x_2) p(x_2, x_3)}{p(x_2)} .
\l{aiugbvv1}
\ee
Equating (\r{aiugbvv}) and (\r{aiugbvv1}), we have
\[
p(x_1, x_3) = \frac{p(x_1, x_2)p(x_3)}{p(x_2)} .
\]
Then
\ba
p(x_1) & = & \sum_{x_3} \frac{p(x_1, x_2)p(x_3)}{p(x_2)} \\
& = & \frac{p(x_1, x_2)}{p(x_2)} ,
\ea
or
\[
p(x_1, x_2) = p(x_1) p(x_2) .
\]
Substituting this into (\r{aiugbvv1}), we have
\[
p(x_1, x_2, x_3) = p(x_1) p(x_2, x_3) ,
\]
i.e., $X_1 \Perp (X_2, X_3)$.  

However, (\r{avuibva}) does not hold in general, because if $X_1 = X_2 = X_3$, we see that 
$X_1 \Perp X_2  \ | \ X_3$ and $X_1 \Perp X_3 \ | X_2$ but $X_1 \not\Perp (X_2, X_3)$.
Note that  $p$ is not strictly positive if $X_1 = X_2 = X_3$.
\l{haha}
\end{eg}

We now present the formulation of an MRF defined on an
arbitrary graph.
Let $G=(V,E)$ be an undirected graph, where
$V = \{1, 2, \cdots, n \}$ is the set of vertices and $E \subset V \times V$ is the set of edges.
We assume that there is no edge in $G$ which joins a vertex to itself.
For any (possibly empty) subset $U$ of $V$,
denote by $G \backslash U$ the graph obtained from $G$ by
removing all the vertices in $U$ and all the edges joining a vertex in $U$.
Let $s(U)$ be the number of components\footnote{
A component of an undirected graph is a subgraph in which any two vertices are connected, 
and which is not connected to any additional vertices in the supergraph.
} in $G \backslash U$.
Denote the sets of vertices of these components by $V_1(U), V_2(U), \cdots,
V_{s(U)}(U)$.
If $s(U) > 1$, we say that $U$ is a cutset in $G$.
Throughout this paper, whenever we remove a subset of vertices $U$ from
$G$, we always assume that we also remove all 
the edges joining a vertex in~$U$.

Consider a collection of random variables $X_i, i \in V$
whose joint distribution is specified by a probability measure $p$ on 
$\calX_1 \times \calX_2 \times \cdots \times \calX_n$, where random
variable $X_i$ is associated with vertex $i$ in graph $G$.
We now define a few Markov properties for random variables $X_1, \cdots,
X_n$ pertaining to a graph $G = (V,E)$:

\begin{defn}[Pairwise Markov Property]
For all distinct $i, j \in V$ such that $\{i, j \} \notin E$, $X_i$ and $X_j$ are independent conditioning on $X_{V - \{ i, j \}}$.
\end{defn}

\begin{defn}[Local Markov Property]
For all $i \in V$, $X_i$ and $X_{V-\ol{N}(i)}$ are independent conditioning on $X_{N(i)}$, where 
$N(i) = \{ j \in V : \{i,j\} \in E \}$ is the set of neighbors of vertex~$i$ and $\ol{N}(i) = N(i) \cup \{i\}$.
\end{defn}

\begin{defn}[Global Markov Property]
Let $\{ U, V_1, V_2 \}$ be a partition of $V$ such that the sets of vertices $V_1$ and 
$V_2$ are disconnected in $G \backslash U$.  Then the sets of random variables 
$X_{V_1}$ and $X_{V_2}$ are independent conditioning on $X_U$.
\l{GMPI}
\end{defn}

\begin{prop}
Random variables $X_i, i \in V$ satisfy the global Markov property if and only if for all cutsets $U$ in $G$, 
the sets of random variables $X_{V_1(U)}, \cdots,
X_{V_{s(U)}(U)}$ are mutually independent conditioning on $X_U$.
\l{GMP2}
\end{prop}

See \cite{YeungLY02} for a proof of Proposition~\r{GMP2}.
When $U = \emptyset$, this proposition states that if the
graph $G$ has more than one
component, i.e., $s(\emptyset) > 1$, then the sets of random variables
$X_{V_1}(\emptyset), \cdots,$ $X_{V_2}(\emptyset), \cdots, X_{V_{s(\emptyset)}}(\emptyset)$ are mutually independent.
Here we regard unconditional mutual independence as a special case of conditional mutual independence.

Denote the Pairwise Markov Property, the Local Markov Property, and the Global Markov Property by 
(P), (L), and (G), respectively.  It can readily be seen from their definitions that (G) $\Rightarrow$ (L) $\Rightarrow$ (P).

\begin{defn}[Markov Random Field]
The probability measure $p$, or equivalently, the
random variables $X_i, i \in V$, are said
to form an MRF represented by a graph $G = (V,E)$ if and only if
the Global Markov Property is satisfied by $X_i, i \in V$.
\label{MRF}
\end{defn}

If $X_i, i \in V$ form an MRF represented by a graph
$G$, we also say that $X_i, i \in V$ form a Markov graph $G$, 
$X_i, i \in V$ are represented by $G$, or $G$ is a (graph) representation for $X_i, i \in V$.  
When $G$ is a path,\footnote{A path is a graph whose vertices can be linearly ordered so that 
every pair of consecutive vertices forms an edge.} we say that $X_i, i \in V$ form a Markov chain.
When $G$ is a tree, we say that $X_i, i \in V$ form a
Markov tree.\footnote{The term ``Markov tree" is also used in the number theory literature in the context of the Markov
number, but it is not to be confused with the Markov tree in this paper.}  
When $G$ is a cycle graph,\footnote{A cycle graph is a graph that consists of a single cycle.} we say that $X_i, i \in V$ form a Markov ring.


In general, $X_i, i \in V$ can be represented by more
than one graph.
In particular,  
$X_i, i \in V$ are always represented by
$K_n$, the complete graph with $n$ vertices.
The graph $K_n$ specifies a degenerate MRF,
because for every $U \subsetneq V$, $U$ is not a cutset in 
$K_n$.
In other words, no Markov constraints are imposed on
$X_i, i \in V$ by $K_n$.

Suppose the random variables $X_i, i \in V$ are represented by both
$G = (V,E)$ and $G' = (V,E')$, where $E' \subsetneq E$, i.e., $G'$ is a proper subgraph of $G$.  Then 
$G'$ imposes a larger set of Markov constraints on $X_i, i \in V$ than
$G$, because a cutset in $G$ is also a cutset in $G'$ (but not vice versa).  Thus 
we are naturally interested in the ``smallest" graph (to be discussed in Section~\r{sec:small}) that represents 
$X_i, i \in V$.

\begin{defn}[Subfield]
A subset of the random variables forming an MRF is called a subfield of the 
MRF.
\end{defn}

\begin{defn}
A $n$-tuple $\bfx = (x_1, x_2, \cdots, x_n) \in \calX_1 \times \calX_2 \times \cdots \times \calX_n$
is called a configuration.
A probability measure $p$ on $\calX_1 \times \calX_2 \times \cdots \times \calX_n$
is strictly positive, denoted by $p > 0$, if 
$p({\bfx}) > 0$
for all configurations $\bfx$.
\end{defn}

If $p > 0$, it can be shown that (G) = (L) = (P) (see for example \cite{Lauritzen96}).  
In general, however, a probability
measure $P$ may contain zero probability masses, i.e., $p(\bfx) = 0$ for some configuration $\bfx$.
For example, if some random variables
in $X_1, X_2, \cdots, X_n$ are functions of other random variables, then $p$ is not strictly positive.

In this paper, we study the structure of MRFs by means of an information-theoretic approach.
Specifically, structural properties of MRFs are obtained
through the investigation of the set-theoretic structure of Shannon's information
measures under the constraints imposed by the MRF.
With this approach, we do not have to manipulate
the underlying probability measure directly.  

An identity involving only Shannon's
information measures (i.e., entropy, mutual information, and their conditional
versions) is referred to as an information identity.
The set-theoretic structure of Shannon's information measures was first studied
in \cite{Hu62}, where it was proved that for every information identity, there is a corresponding  
set identity.
This was further developed into the theory of $I$-Measure in \cite{Yeung91}.
Under this framework, every Shannon's information measure can formally be regarded as the value
of a unique signed measure called the $I$-Measure, denoted by $\mu^*$, on a set corresponding to
that Shannon's information measure.
This establishes a complete set-theoretic interpretation of Shannon's information measures.

Subsequent to \cite{Yeung91}, the structure of the 
$I$-Measure for a Markov chain and more generally an MRF was investigated in \cite{KawabataY92}
and \cite{YeungLY02}, respectively.
In particular, it was proved in \cite{KawabataY92} that the $I$-Measure for a Markov chain is always nonnegative,
and an information diagram that displays the special structure of the $I$-Measure for a Markov chain was obtained.

The current work, consisting of the following three main results, is built on \cite{Hu62, Yeung91, KawabataY92, YeungLY02}:
\begin{enumerate}
\item
Let $X_i, i \in V$ be any set of random variables that form an MRF represented by a graph $G$, and 
let $X_i, i \in V'$, where $V' \subset V$, be any subfield of the MRF.
We determine the smallest graph $G^*(V')$ that can always represent $X_i, i \in V'$. 
\item
The $I$-Measure of an MRF is always nonnegative if and only if the MRF is represented
by either a path or a forest of paths.\footnote{A forest of paths is a graph with at least two components such that each component is a path.}
\item
We develop a recursive approach for constructing 
an information diagram that displays the special structure of the $I$-Measure for an MRF.

\end{enumerate}
The rest of the paper is organized as follows.
Section~\r{sec2} contains an overview of the concepts and tools to be used in
this paper.  
In Section~\r{sec3}, we define the graph $G^*(V')$ and establish that $G^*(V')$ is 
the smallest graph that can always represent
the subfield $X_i, i \in V'$.
Applying this result to Markov trees, we obtain in 
Section~\r{sec4} a necessary and sufficient condition for a subfield of a Markov tree
to form a Markov subtree.
In Section~\r{sec5}, we establish that Markov chains are essentially the 
only MRFs  for which the $I$-Measure is always nonnegative.
In Section~\r{sec6}, we develop a recursive approach for constructing 
an information diagram that displays the special structure of the $I$-Measure for an MRF.
The paper is concluded in Section~\r{sec7}.

\section{Preliminaries}
\l{sec2}
In this section, we introduce the notations and present the preliminaries for the 
rest of the paper.  For a detailed discussion, we refer the readers to \cite[Chapters~3 and 12]{Yeung08} and the references therein.

\subsection{$I$-Measure}
\l{sec2.1}
We first give an
overview of the basics of the $I$-Measure.
Let $X_i, i \in V = \{ 1, 2, \cdots, n \}$ be jointly distributed discrete random
variables, and $\tilde{X}$ be a set variable corresponding to a random
variable $X$.
We note that the $I$-Measure does not have to be defined in the context of
an MRF, but here we use $V$ (the vertex set of a graph) 
as the index set of the random variables for the sake of convenience.
Here we assume that $H(X_i) < \infty$ for $1 \leq i \leq n$,
so that the $I$-Measure \cite{Yeung91}
for $p$ is well-defined.

Define the universal set $\Omega_V$ to be $\bigcup_{i \in V} \tilde{X}_i$ and let $\calF_V$
be the $\sigma$-field generated by $\{ \tilde{X}_i, i \in V \}$.  The atoms
of $\calF_V$ have the form $\bigcap_{i \in V} Y_i$, where $Y_i$ is either $\tilde{X}_i$ or
$\tilde{X}_i^c$.  Let ${\calA_V} \subset {\calF_V}$ be the set of all the atoms of
$\calF_V$ except for $\bigcap_{i \in V} \tilde{X}_i^c$, which is equal
to the empty set because
\[
\bigcap_{i \in V} \tilde{X}_i^c = \left( \bigcup_{i \in V} \tilde{X}_i 
\right)^c = (\Omega_V)^c = \emptyset .
\]
Note that $|{\calA_V}| = 2^n-1$.
In the rest of the paper, when we refer to an atom of $\calF_V$, we always
mean an atom in $\calA_V$ unless otherwise specified.

To simplify notation, we will use $X_U$ to denote $(X_i,i \in U)$ and
$\tilde{X}_U$ to denote $\bigcup_{i \in U} \tilde{X}_i$ for any $U \subset V$.
We will not distinguish between $i$ and the singleton containing $i$.
It was shown in \cite{Yeung91}
that there exists a unique {\em signed} measure $\mu^*$ on $\calF_V$
which is consistent with all Shannon's information measures via the following
substitution of symbols:
\begin{eqnarray*}
H/I & \rightarrow & \mu^* \\
, & \rightarrow & \cup \\
; & \rightarrow & \cap \\
| & \rightarrow & - 
\end{eqnarray*}
where ``$-$" denotes the set difference.  For example, 
\[
\mu^*( (\tX_1 \cup \tX_2) \cap \tX_3 - \tX_4 ) = I( X_1, X_2; X_3 | X_4 ) .
\]
For all $A \in \calA_V$, $\mu^*(A)$ is a linear combination of $H(X_B)$ for nonempty subsets $B$ of $V$.

Note that $\mu^*$ in general is not nonnegative.  However, if $X_i, i \in V$ form a Markov chain, 
then $\mu^*$ is always nonnegative
\cite{KawabataY92}.  See Section~\r{sec5} for further discussions.

\subsection{Full Conditional Mutual Independency}
\l{sec:FCMI}
\begin{defn}
Let $\{ T, Q_1, Q_2, \cdots, Q_k \}$ be a partition of $V'$, where $k \ge 2$ and $V' \subset V$.  The tuple 
$K=(T; Q_i, 1 \le i \le k)$ defines
the following conditional mutual independency (CMI) on $X_i, i \in V$:
\[ 
X_{Q_1}, X_{Q_2}, \cdots, X_{Q_k} 
\ \mbox{are mutually independent conditioning on $X_T$.}
\]
If $V' = V$, $K$ is called a full conditional mutual independency (FCMI).
\end{defn}

\begin{eg}
For $n = 6$, $K = ( \{4\}; \{1,3\}, \{2,5\}, \{6\})$ defines the FCMI
\[
(X_1,X_3), (X_2,X_5), X_6 \ \mbox{are mutually independent conditioning on $X_4$.}
\]
However, for $n = 7$, $K$ is not an FCMI because 
$\{ \{4\}, \{1,3\},$ $ \{2,5\}, \{6\} \}$ is not a partition of $\{ 1, 2, \cdots, 7 \}$.
\end{eg}

\begin{defn}
Let $K= (T; Q_i, 1 \leq i \leq k)$ be an FCMI on $X_i, i \in V$.  The image of $K$,
denoted by ${\it Im} (K)$, is the set of atoms of $\calF_V$ of the form
\be
\left( \bigcap_{i=1}^k \bigcap_{j \in W_i} \tilde{X}_j \right) - 
\tilde{X}_{T \cup ( \bigcup_{i=1}^k (Q_i - W_i) ) } 
\l{atom}
\ee
where $W_i \subset Q_i$, $1 \leq i \leq k$,
and there exist at least two $i$ such that $W_i \neq \emptyset$.
\l{def_atom}
\end{defn}

The following proposition gives a more explicit expression for $\Im(K)$.
The proof is elementary and so is omitted. 

\begin{prop}
\l{prop:atom}
Let $K= (T; Q_i, 1 \leq i \leq k)$ be an FCMI on $X_i, i \in V$.  Then
\[
{\it Im} (K) =
\left\{ A \in \calA_n : A \subset \bigcup_{1 \le i < j \le k} 
(\tilde{X}_{Q_i} \cap \tilde{X}_{Q_j} - \tilde{X}_T) \right\}.
\]
\end{prop}

In the rest of paper, we denote the atom $\tX_1 \cap \tX_2 \cap \tX_3^c$ of $\calF_{\{1,2,3\}}$ by
$12\bar{3}$, etc.

\begin{eg}
Let $n=3$ and consider the FCMI
$K = ( \emptyset; \{1\}, \{2\}, \{3\} )$.
Then $\Im(K)$ is the set containing all the atoms in 
$(\tX_1 \cap \tX_2) \cup (\tX_1 \cap \tX_3) \cup (\tX_2 \cap \tX_3)$, as given 
by Proposition~\r{prop:atom}.
This is illustrated in Fig.~\r{figeg2}.  Equivalently, the atoms in $\Im(K)$
are $1 2 \bar{3}$, $1 \bar{2} 3$, $\bar{1} 2 3$, and $123$, as given by Definition~\r{def_atom}.
For example, for the atom $1 2 \bar{3}$, we have $W_1 = \{1\}$, $W_2 = \{2\}$,
and $W_3 = \emptyset$, so there are at least two $i$ such that $W_i \ne \emptyset$.
\l{eg2}
\end{eg}

The following theorem from \cite{Yeung08} will be useful for proving some of the results in this work.

\begin{thm}
Let $K$ be an FCMI on $X_i, i \in V$. Then $K$ 
holds if and only if 
$\mu^*(A) = 0$ for all $A \in {\it Im} (K)$.
\l{full_tatom}
\end{thm}

Thus the effect of an FCMI $K$ on the joint probability distribution of $X_i, i \in V$
is completely characterized by $\Im(K)$.
We remark that if $\{ T, Q_1, Q_2, \cdots, Q_k \}$ is a partition of $V'$ where $V' \subsetneq V$, then
$K = (T; Q_1, Q_2, \cdots, Q_k)$ holds if and only if $\mu^*$ vanishes on all the sets prescribed in 
(\r{atom}), although these sets are no longer atoms of $\calF_V$.

\begin{eg}
Following Example~\r{eg2}, the random variables $X_1, X_2$, and $X_3$ are mutually independent
if and only if $\mu^*$ vanishes on the atoms $1 2 \bar{3}$, $1 \bar{2} 3$, $\bar{1} 2 3$, and $123$.
\end{eg}

Let $A = \bigcap_{i \in V} \tilde{Y}_i$ be a nonempty atom of $\calF_V$.
Define the set
\be
U_A = \{ i \in V :  \tilde{Y}_i = \tilde{X}_i^c \}.
\l{U_Adef}
\ee
Note that $A$ is uniquely specified by $U_A$ because
\[
A = \left( \bigcap_{i \in V - U_A} \tilde{X}_i \right) \cap
\left( \bigcap_{i \in U_A} \tilde{X}_i^c \right)
= \left( \bigcap_{i \in V - U_A} \tilde{X}_i \right) \cap \left( \bigcup_{i \in U_A} \tilde{X}_i \right)^c
= \left( \bigcap_{i \in V - U_A} \tilde{X}_i \right) - \tilde{X}_{U_A}.
\]
Also note that in the definition of $U_A$, its dependence on $V$ is implicit, and what the set $V$ is should be clear from
the context.

Define $w(A) = n - |U_A|$ as the 
{\em weight} of the atom $A$, 
the number of $\tilde{X}_i$ in $A$ which are not complemented.
We now show that an FCMI $K=(T; Q_i, 1 \le i \le k)$ is uniquely
specified by ${\it Im}(K)$.  First, by letting $W_i = Q_i$ for $1 \leq i \leq k$
in (\r{atom}), we see that the atom
\[
\left( \bigcap_{j \in \bigcup_{i=1}^k Q_i} \tilde{X}_j \right) - \tilde{X}_T 
\]
is in ${\it Im}(K)$, and it is the unique atom in ${\it Im}(K)$ with the largest weight.
From this atom, $T$ can be determined.  To determine $Q_i, 1 \le i \le k$, 
we define
a relation $q$ on $T^c = V - T$ as follows.  For $l, l' \in T^c$,
$(l,l')$ is in $q$ if and only if one of the following is satisified:
\begin{list}%
{\roman{cond})}{\usecounter{cond}}
\item
$l=l'$;
\item
$l \ne l'$ and the atom
\be
\tilde{X}_l \cap \tilde{X}_{l'} \cap \left( \bigcap_{j \in V - \{ l, l' \}}
  \tilde{X}_j^c \right)
\l{89nv45}
\ee
is not in ${\it Im}(K)$.  
\end{list}
The idea of ii) is that $(l,l')$ is in $q$ 
if and only if $l, l' \in Q_i$ for some
$1 \leq i \leq k$, which can be seen as follows.
If $l, l' \in Q_i$ for some $i$, then the atom in (\r{89nv45}) is not
in $\Im(K)$ by Definition~\r{def_atom} because $\{l, l' \} \subset W_i$ and so 
$W_i \ne \emptyset$ but $W_j = \emptyset$ for all $j \ne i$
(an atom in $\Im(K)$ has
at least two $i$ such that $W_i \ne \emptyset$).
On the other hand, if $l \in Q_i$ and $l' \in Q_{i'}$ where $i \ne i'$,
then by letting $W_i = \{ l \}$ and $W_{i'} = \{ l' \}$, we see that 
the atom in (\r{89nv45}) is in $\Im(K)$.

Then $q$ is reflexive by i), and is
symmetric because the definition of $q$ is symmetrical in $l$ and $l'$.  Moreover, $q$ is transitive from the discussion above
because if $l, l' \in Q_i$ for some $1 \le i \le k$ and $l', l^{\prime\prime} \in Q_{i'}$ for some $1 \le i' \le k$,
then $i = i'$ and $l, l^{\prime\prime} \in Q_i$.  In other words, $q$ is an
{\em equivalence relation}
that partitions $T^c$ into $\{ Q_i, 1 \le i \le k \}$.
Therefore, $K$ can be recovered from ${\it Im}(K)$, and so it is uniquely specified by ${\it Im}(K)$.

Let $\Pi = \{ K_l, 1 \le l \le m \}$ be a collection of FCMIs on $X_i , i \in V$, and define
\[
\Im(\Pi) = \bigcup_{l=1}^k \Im(K_l) .
\]
Since $\Pi$ holds if and only if $K_l$ holds for all $l$, it follows from Theorem~\r{full_tatom} that
$\Pi$ holds if and only if $\mu^*(A) = 0$ for all $A \in \Im(\Pi)$.  
Thus the effect of a collection $\Pi$ of FCMIs on the joint probability distribution
of $X_i, i \in V$ is completely characterized by $\Im(\Pi)$.

Consequently, for two collections $\Pi_1$ and $\Pi_2$ of FCMIs, $\Pi_1 \Rightarrow \Pi_2$
if and only $\Im(\Pi_1) \supset \Im(\Pi_2)$, and $\Pi_1 = \Pi_2$ 
if and only $\Im(\Pi_1) = \Im(\Pi_2)$.

One can interpret $\Im(K)$ as the ``footprint" of an FCMI $K$. Then the footprint 
of a collection $\Pi$ of FCMIs is simply the union of the footprints of the individual
FCMIs in $\Pi$.  However, two different collections of FCMIs may have the same footprints,
as shown in the next example. 
Thus unlike an FCMI, a collection of FCMIs is in general not uniquely specified by its image.

\begin{eg}
Let $n=3$.  Let $\Pi_1 = \{ K_1 \}$ and \ $\Pi_2 = \{ K_2 , K_3 \}$, where
\ba
K_1 & = & ( \emptyset ; \{1\}, \{2\}, \{3\} ) \\
K_2 & = & ( \emptyset ; \{1, 2\}, \{3\} ) \\
K_3 & = & ( \{3\}; \{1\}, \{2\} ) .
\ea
Then $\Pi_1 \ne \Pi_2$ but $\Im(\Pi_1) = \Im(\Pi_2)$.
\l{qv5gea}
\end{eg}

It was shown in \cite{Malvestuto92, GeigerP93} that full conditional (mutual) independence are
axiomatizable.  This can be regarded as an alternative characterization of FCMIs,
which however is not in closed form.

\subsection{Markov Random Field}
\l{sec:MRF}
In the definition of an MRF, 
each cutset $U$ in $G$ specifies an FCMI on 
$X_1, X_2, \cdots, X_n$, denoted by $[U]$.  Formally,
\ba
[U] : & & \hspace{-.2in} X_{V_1(U)}, \cdots,
X_{V_{s(U)}(U)} \ \mbox{are mutually independent}  \\
 & &  \hspace{-.2in} \mbox{conditioning on $X_U$.} 
\ea
Then in light of (\r{U_Adef}), for $A \in \calA_n$ such that $s(U_A) > 1$,
$[U_A]$ is the FCMI induced by the cutset $U_A$.
It follows that 
$X_1, X_2, \cdots, X_n$ form a Markov graph $G$ 
if and only if
\be
\bigwedge_{A \in \calA_n : s(U_A) > 1} [U_A] \stackrel{\triangle}{=} [U_G],
\ee
where `$\wedge$' denotes `logical AND'.
This is the collection of FCMIs induced by graph $G$. 

\begin{defn}
Let $G = (V, E)$ be a graph.  For an atom $A$ of $\calF_V$, if $s(U_A)=1$, i.e., $G \bs U_A$
is connected, then $A$ is a Type~I atom of $G$, otherwise, i.e., $s(U_A) > 1$, $A$ is a Type~II atom of $G$.  The sets
of all Type~I and Type~II atoms of $G$ are denoted by $\calT_{{\rm I}}(G)$ and $\calT_{{\rm II}}(G)$, respectively.
\l{type}
\end{defn}


\begin{defn}
For a graph $G = (V,E)$, the image of $G$ is defined by
\[ 
\Im(G) = \Im ( [U_G ] ) .
\]
\end{defn}

\begin{thm} (cf. \cite[Theorem~12.25]{Yeung08}) 
$\Im(G) = \calT_{{\rm II}}(G)$.
\l{qarocf}
\end{thm}

The above theorem gives a precise characterization of $\Im(G)$.  It follows from the discussion in Section~\r{sec:FCMI} that 
$X_1, X_2, \cdots, X_n$ form a Markov graph $G$ if and only if $\mu^*(A) = 0$ for all $A \in  \calT_{{\rm II}}(G)$,
i.e., $\mu^*$ vanishes on all the Type~II atoms of $G$.

\begin{eg}
For the cycle graph $G$ in Fig.~\r{figeg5}, $\calT_{{\rm II}}(G) = \{ 1\bar{2}3\bar{4}, \bar{1}2\bar{3}4 \}$.  Random variables $X_1, X_2, X_3$ 
and $X_4$
are represented by $G$ if and only if $\mu^*(1\bar{2}3\bar{4}) = \mu^*(\bar{1}2\bar{3}4) = 0$.
\l{eg5}
\end{eg}

A graph $G = (V, E)$ and the collection $[U_G]$ of FCMIs it induces uniquely specify each other, because for distinct
$u, v \in V$, $\{ u, v \} \in E$ if and only if the FCMI $(V- \{u,v\}; \{u\}, \{v\})$ is not in 
$[ U_G ]$.  This can be seen as follows.
If $\{ u, v \} \in E$, then $V - \{ u, v \}$ is not a cutset in $G$, and so $(V- \{u,v\}; \{u\}, \{v\}) \not\in [ U_G ]$.  On the other hand,
if $(V- \{u,v\}; \{u\}, \{v\}) \not\in [ U_G ]$, then $V - \{ u, v \}$ is not a cutset in $G$, which implies $\{ u, v \} \in E$.

Although a collection of FCMIs is in general not uniquely determined by its
image (cf. Example~\r{qv5gea}), the following proposition asserts that a graph $G$ (and hence $[U_G]$) is uniquely determined by its image
$\Im(G)$.

\begin{prop}
For a graph $G = (V,E)$, $\{ u, v \} \in E$ if and only if the atom 
\be
\tX_u \cap \tX_v - \tX_{V - \{ u, v \} }
\l{4c9ayr}
\ee
is not in $\Im(G)$.
\end{prop}

\pf
Denote the atom in (\r{4c9ayr}) by $A$.  If $\{ u, v \} \not\in E$, then $G$ induces the FCMI $K = (V - \{ u, v \}; \{ u \}, \{ v \})$.
Obviously, $\Im(K) = \{A\}$, and hence $A \in 
\Im(K) \subset \Im(G)$.

To prove the converse, assume that atom A is in $\Im(G)$, and 
specifically in some $\Im([U_{A'}])$ such that $s(U_{A'}) > 1$.  It follows from Definition~\r{def_atom}
that in order for $A$ to be in
$\Im([U_{A'}])$, it is necessary for $u$ and $v$ to be in different sets in 
$V_1(U_{A'}), V_2(U_{A'}), \cdots,$ $V_{s(U_{A'})}(U_{A'})$.
This implies that $V - \{ u, v \}$ is a cutset in $G$,
and hence $\{ u, v \} \not\in E$. 
\endpf

\bigskip
With this proposition, a graph $G$ can be recovered from $\Im(G)$ as follows.
Start with the complete graph $K_n$.  If there exists an atom in $\Im(G)$ as prescribed by 
(\r{4c9ayr}) for some distinct $u, v \in V$, then remove edge $\{u, v \}$ from the graph.
Repeat this step until no more edges can be removed.  
Note that this algorithm produces a unique graph, i.e., $G$.
As a corollary, the uniqueness of the Markov graph induced by $\Im(G)$ is proved, i.e., 
for two graphs $G = (V,E)$ and $G' = (V,E')$ where $E \ne E'$,
$\Im(G) \ne \Im(G')$.


\subsection{Smallest Graph Representation}
\l{sec:small}
As discussed in Section~\r{sec:intro}, 
we are interested in the ``smallest" graph that can represent a given set of random variables $X_i, i \in V$.
To fix ideas, we first give a formal definition of this notion.

\begin{defn}
A graph $G = (V, E)$ is the smallest graph representation for a set of random variables $X_i, i \in V$ if
$G$ is a representation for $X_i, i \in V$ and is a subgraph 
of any representation $G'$ for $X_i, i \in V$.
\end{defn}

We know from Section~\r{sec:MRF} that a graph $G = (V,E)$ can represent $X_i, i \in V$ if and only 
if $\mu^*$ vanishes on all the atoms in $\Im(G)$.  
Note that if $G$ is a subgraph of $G'$, then a cutset in $G'$ is also a cutset in $G$.  
It follows that $[U_{G'}]$ is a ``subset" of $[U_{G}]$, and hence 
$\Im(G') \subset \Im(G)$.

For the given set of random variables $X_i, i \in V$, let $\calA_{\rm II}$ be the set of nonempty atoms of $\calF_V$ 
on which $\mu^*$ vanishes.
Following the last paragraph, if $G$ is the smallest representation for $X_i, i \in V$, then
$\Im(G) \subset \calA_{\rm II}$
and $\Im(G') \subset \Im(G)$ for any representation $G'$ for $X_i, i \in V$.
The next theorem gives a characterization of such a graph
if it exists.

\begin{thm}
For a given set of random
variables $X_i, i \in V$, let $\calA_{\rm II}$ be the set of nonempty atoms of $\calF_V$ 
on which $\mu^*$ vanishes.
Let $\hat{G} = (V,\hat{E})$ be such that 
$\{ u, v \} \in \hat{E}$ if and only if the atom in (\r{4c9ayr})
is not in $\calA_{\rm II}$.  Then if the smallest graph representation for $X_i, i \in V$ exists,
it is equal to $\hat{G}$.
\l{alrya944}
\end{thm}

We first prove the following two lemmas.

\begin{lemma}
Every graph that can represent $X_i, i \in V$ contains $\hat{G}$ as a subgraph.
\l{lema}
\end{lemma}

\pf
Let $G' = (V,E')$ be any graph that can represent $X_i, i \in V$.
Consider any edge $\{u,v\}$ in $\hat{G}$, i.e., $\{u,v\} \in \hat{E}$.  By construction, the atom in (\r{4c9ayr}) is not in $\calA_{\rm II}$. 
Then $\{ u, v \} \in E'$,
otherwise the FCMI $( V- \{ u, v \}; \{ u \}, \{ v \} )$ holds, i.e.,
\[
I(X_u;X_v|X_{V- \{u,v\}}) = \mu^*(\tX_u \cap \tX_v - \tX_{V - \{ u, v \} })  = 0 ,
\]
which is a contradiction because the atom in (\r{4c9ayr}) is not in $\calA_{\rm II}$.
Thus if $G'$ can represent $X_i, i \in V$, then $G'$ contains $\hat{G}$ as a subgraph.
\endpf

\begin{lemma}
If $\{ u, v \}$ is an edge in every graph that can represent $X_i, i \in V$,
then $\{ u, v \}$ is an edge in $\hat{G}$.
\l{lemb}
\end{lemma}

\pf
Let $\{ u, v \}$ be an edge in every graph that can represent $X_i, i \in V$.  If a graph 
does not contain $\{ u, v \}$, then it cannot represent $X_i, i \in V$.
In particular, the graph $K_n \bs \{u,v\}$ obtained by removing $\{ u, v \}$ from the complete graph 
$K_n$ cannot represent $X_i, i \in V$.  Since the only FCMI imposed by $K_n \bs \{u,v\}$
is $[ \{u,v\} ]$ (i.e., $X_u$ and $X_v$ are independent conditioning on $X_{V - \{u,v\}}$), this means
that $X_u$ and $X_v$ are not independent conditioning on $X_{V - \{u,v\}}$, or 
$\mu^*(\tX_u \cap \tX_v - \tX_{V-\{u,v\}}) > 0$.  In other words, the atom $\tX_u \cap \tX_v - \tX_{V-\{u,v\}}$
is not in $\AII$, which implies that $\{ u, v \}$ is an edge in $\hat{G}$.
\endpf

\bigskip
\noindent
{\bf Proof of Theorem~\r{alrya944}}
Assume the smallest graph representation for $X_i, i \in V$ exists and let it be $\tG$.
By Lemma~\r{lema}, $\hat{G}$ is a subgraph of $\tG$.  On the other hand, since
$\tG$ is a subgraph of every graph that can represent $X_i, i \in V$, Lemma~\r{lemb} 
implies that $\tG$ is a subgraph of $\hat{G}$.  Hence, $\tG = \hat{G}$.
\endpf

\begin{cor}
The smallest graph representation for $X_i, i \in V$ exists if and only if $\hat{G}$ 
is a representation for $X_i, i \in V$.
\l{corr}
\end{cor}

\pf
Assume that the smallest graph representation for $X_i, i \in V$ exists.  By Theorem~\r{alrya944},
it is equal to $\hat{G}$, and so $\hat{G}$ 
is a representation for $X_i, i \in V$.  Conversely, if $\hat{G}$ 
is a representation for $X_i, i \in V$, then by Lemma~\r{lema}, it is the smallest representation for 
$X_i, i \in V$.
\endpf

\begin{eg}
Let $n=3$ and consider $\mu^*$ such that
\be
\AII = \{ 12\bar{3}, 1\bar{2}3 \}.
\l{farwioh}
\ee
Accordingly,
the graph $\hat{G}$ defined in Proposition~\r{alrya944} is illustrated in Fig.~\r{figeg6}.  However,
\[
\Im(\hat{G}) = \{ 12\bar{3}, 1\bar{2}3, 123 \} \not\subset \AII,
\]
 i.e., $\hat{G}$ cannot represent $X_1, X_2$, and $X_3$.
Then by Corollary~\r{corr}, there does not exist a smallest graph representation of
$X_1, X_2$, and $X_3$.
\l{eg6}
\end{eg}

The above example shows that the smallest graph representation may  not exist for a given set of
random variables.
However, if $\AII = \Im(G)$
for some graph $G$, then $G$ is in fact the smallest graph representation for $X_i, i \in V$.
This is proved in the next proposition.

\begin{prop}
If $\AII = \Im(G)$
for some graph $G$, then $G$ is the smallest graph representation for $X_i, i \in V$.
\end{prop}

\pf
We see that a graph $G$ can be recovered from its image 
$\Im(G)$ using the algorithm described at the end of Section~\r{sec:MRF}, and in fact
$G = \hat{G}$.  Therefore, $\Im(\hat{G}) = \Im(G) = \AII$, which implies that
$\Im(\hat{G}) \subset \AII$.  Hence, $\hat{G}$ is a graph representation for $X_i, i \in V$.
It then follows from Theorem~\r{alrya944} that $\hat{G}$, i.e., $G$, is the smallest graph representation for
$X_i, i \in V$.
\endpf


\bigskip
\noindent
To our knowledge, Corollary~\r{corr} is new.  A related result can be found in \cite{PearlP85},
where it was proved that if the underlying probability measure $p$ is strictly positive, then
the smallest graph representation for $X_i, i \in V$ always exists and is equal to $\hat{G}$.

\begin{eg}
In Example~\r{eg6}, the constraint (\r{farwioh}) is equivalent to
$X_1 \Perp X_2 \ | \ X_3$ and $X_1 \Perp X_3 \ | \ X_2$, while 
$
[U_{\hat{G}}] = [U_{\emptyset}, U_{\{2\}}, U_{\{3\}}]
$
consists of the FCMIs
\[
X_1 \Perp (X_2,X_3), \ X_1 \Perp X_3 \ | \ X_2, \ X_1 \Perp X_2 \ | \ X_3.
\]
We have shown in Example~\r{haha} that 
\[
\left.
\begin{array}{l}
X_1 \Perp X_2 \ | \ X_3 \\
X_1 \Perp X_3 \ | \ X_2
\end{array}
\right\} \ \Rightarrow \ X_1 \Perp (X_2,X_3) ,
\]
holds if the underlying 
probability distribution $p$ is strictly positive, or $p > 0$, but does not hold in general.
This means that if $p > 0$, then $\hat{G}$ represents $X_1, X_2$, and $X_3$, 
but in general it does not.  These conclusions are consistent with the
result in \cite{PearlP85} and the discussion in Example~\r{eg6}, respectively.
\end{eg}

\section{Subfield of a Markov Random Field}
\l{sec3}
Let $X_i, i \in V$ form an MRF represented by some graph $G = (V,E)$.  Note that such a graph $G$ 
can always be found, because $K_n$ is always a representation of $X_i, i \in V$.
Let $V'$ be a subset of $V$.  In this section, we seek the smallest graph that can always represent 
the subfield $X_i, i \in V'$.

\begin{defn}
Let $G= (V,E)$ and $G' = (V',E')$ where $V' \subset V$.  If $[U_G] \Rightarrow [U_{G'}]$,
we write $G \Rightarrow G'$.
\end{defn}


Let $X_i, i \in V$ form an MRF represented by a graph $G$.  Following the definition above,
if $G \Rightarrow G'$, then $X_i, i \in V'$ form
an MRF represented by $G'$.

\begin{defn}
Let $G= (V,E)$, and let $V' \subset V$.  Let $G^*(V') = (V',E')$ be such that for distinct $u, v \in V'$, 
$\{ u,v \} \in E'$ if and only if there exists a path between $u$ and $v$ in $G$ on which all the intermediate vertices 
are in $V - V'$.
\l{G*}
\end{defn}

Obviously, $G^*(V) = G$.
We will prove in Theorem~\r{quaff;ova}, the main theorem of this section, that $G^*(V')$ is the smallest $G'$ such 
that $G \Rightarrow G'$.  

\begin{eg}
Consider an MRF represented by the graph $G$ in Fig.~\r{figeg7}, which indeed is a Markov chain.
Let $V' = \{ 1, 3, 5, 6 \}$.  Then $G^*(V')$ is illustrated as the overlay graph in grey.
\l{eg7}
\end{eg}

\begin{eg}
Consider an MRF represented by the more elaborate graph $G$ in Fig.~\r{figeg8}.
Let $V' = \{ 1, 2, 5, 6, 8, 9 \}$.  Then $G^*(V')$ is illustrated as the overlay graph in grey.
\l{eg8}
\end{eg}

Consider $V^\" \subset V' \subset V$.  The next proposition asserts that $G^*(V^\")$ can be obtained in two steps.  
First obtain $G^*(V')$ from $G$
by applying Definition~\r{G*}.  Then obtain $G^*(V^\")$ from $G^*(V')$ by applying Definition~\r{G*}
again with $G^*(V')$ in place of $G$.

\begin{prop}
Let $G = (V, E)$ and $V^\" \subset V' \subset V$.  Then $G^*(V^\") = (G^*(V'))^*(V^\")$.
\l{V"}
\end{prop}

\pf
See Appendix~\r{**}.

\bigskip
Consider $G \bs (V-V') = (V',E^{\prime\prime})$, where
\[
E^{\prime\prime} = \{ \{ v, w \} : v, w \in V' \ \mbox{and} \ \{ v, w \} \in E \} .
\]
For distinct $v, w \in V'$, 
if $\{v,w\} \in E^{\prime\prime}$, then $\{v,w\} \in E'$ by the definition of $G^*(V')$.
In other words, $G^*(V')$ always contains $G \bs (V-V')$ as a subgraph.
However, $G^*(V') \ne G \bs (V - V')$ in general.  In other words, $G^*(V')$ is not necessarily a subgraph of $G$.
The following proposition gives the condition for $G^*(V')$ to be exactly equal to $G \bs (V-V')$.

\begin{prop}
Let $G= (V,E)$, and let $V' \subset V$.  Let $\rho (V')$ be the set of elements of $V'$ such that some of their neighbors are
in $V-V'$, i.e., 
\be
\rho (V') = \{ v \in V' : \{u,v\} \in E \mbox{ for some $u \in V - V'$} \}.
\l{aiognv}
\ee
Then $G^*(V ') = G \bs (V - V')$ if and only if for distinct $v, w \in \rho(V')$, if $\{v,w\}$ is not an edge in 
$G \bs ( V - V' )$, then there exists no path between $v$ and $w$ in $G$ on which all the vertices other than
$v$ and $w$ are in $V-V'$. 
\l{ar5hvva}
\end{prop}

\pf
Note that $G^*(V ') = G \bs (V - V')$ is equivalent to $E' = E^{\prime\prime}$.
We already have proved that $E^{\prime\prime} \subset E'$ always holds, so we only need to prove that 
the condition in the proposition
for $G^*(V ') = G \bs (V - V')$ is necessary and sufficient for $E' \subset E^{\prime\prime}$.

For any distinct $v, w, \in V'$, consider two cases.  If either $v$ or $w$ is not in $\rho(V')$, then $\{ v, w \} \in E'$ 
implies $\{ v, w \} \in E^{\prime\prime}$.  If both $v$ and $w$ are in $\rho(V')$, then the condition in the proposition
for $G^*(V ') = G \bs (V - V')$ is necessary and sufficient for $\{ v, w \} \in E'$ to imply $\{ v, w \} \in E^{\prime\prime}$.
The proposition is proved.
\l{quiff;c}
\endpf


\begin{eg}
Consider the graph $G$ in Fig.~\r{figeg9} and let $V' = \{ 2, 3, 4 \}$.
Here $\rho(V') = V'$ because each vertex in $V'$ is connected to some vertex
in $V - V'$.  Now $\{ 2, 4 \}$ is the only pair of vertices that is not an edge in $G \bs (V-V')$.  
Since there exists no path between vertices 2
and 4 on which all the vertices other than 2 and 4 are in $V-V' = \{1, 5\}$, by Proposition~\r{quiff;c},
$G^*(V ') = G \bs (V - V')$, which is illustrated as the overlay graph in grey.
\l{eg9}
\end{eg}

\begin{cor}
Let $V = \{ 1, 2, \cdots, n \}$ and $V' = V - \{ n\}$, where $n \ge 2$.  
Let $X_i, i \in V$ be represented by a graph $G = (V,E)$ such that 
$\{ n-1, n \} \in E$ and $n-1$ is the only neighbor of $n$.  Then 
$X_i, i \in V'$ is represented by $G \bs \{ n\}$.
\l{q5;once}
\end{cor}

\pf
This is a special case of Proposition~\r{ar5hvva} with $\rho(V') = \{ n-1 \}$.
\endpf 

\bigskip
\noindent
As discussed above, $G^*(V')$ always contains $G \bs (V-V')$ as a subgraph.
The next theorem gives an alternative characterization of $G^*(V')$ that describes
the relation between $G^*(V')$ and $G \bs (V-V')$ more explicitly.
For $U \subset V$, let
\[
\phi(U) = \{ v \in V - U : \{v,w\} \in E \ \mbox{for some $w \in U$} \}
\]
be the set of neighbors of $U$ in graph $G$,\footnote{Note that $\phi(U) = \rho(V-U)$, where
$\rho$ is defined in (\r{aiognv}).}
and 
\[
\kappa(U) = \{ \{ u,v \} : u, v \in U \} 
\]
be the set of edges of the clique formed by the vertices in $U$.

\begin{thm}
Let $G= (V,E)$.  For $V' \subset V$, let $G^*(V') = (V',E')$ and $G \bs (V-V') = (V', E^\")$.
Then
\be
E' = E^\" \cup \bigcup_{i=1}^{s(V')}  \kappa( \phi (V_i(V')) ,
\l{qiunvaf}
\ee
where $V_1(V'), V_2(V'), \cdots, V_{s(V')}(V')$ are the components of $G \bs V'$.
\l{renrvf}
\end{thm}

\pf
To facilitate our discussion, let $\tE$ denote the set on the right hand side of (\r{qiunvaf}).
We first prove that $E' \subset \tE$.  By Definition~\r{G*}, if $\{u,v\} \in E'$, then
there exists a path between $u$ and $v$ in $G$ on which all the intermediate vertices
are in $V - V'$.
Denote this set of vertices in $V-V'$ by $S'$.  If $S' = \emptyset$,
then we have $\{u,v\} \in E^\"$.  Otherwise, since the vertices in $S'$ are connected in 
$G \bs V'$, $S'$ is a subset of $V_i(V')$ for some $1 \le i \le s(V')$.  As such, $u, v  \in \phi (V_i(V'))$
and hence $\{u,v\} \in \kappa( \phi (V_i(V'))$.  This completes the proof for $E' \subset \tE$.

It remains to prove that $\tE \subset E'$.  Let $\{u,v\} \in \tE$.  If $\{u,v\} \in E^\"$, then 
$\{u,v\} \in E'$ because $E^\" \subset E'$ as discussed.  If
$\{u,v\} \in \kappa( \phi (V_i(V'))$ for for some $1 \le i \le s(V')$, then
$u, v \in \phi (V_i(V'))$, i.e., there exists
$u', v' \in V_i(V')$ ($u'$ and $v'$ are not necessarily distinct) such that $\{u,u'\}, \{v,v'\}
\in E$.  Since $u'$ and $v'$ are in the same component of $G \bs V'$, namely
$V_i(V')$, they are connected and it follows that there exists a path between $u$ 
and $v$ in $G$ on which all the intermediate vertices
are in $V - V'$.  Therefore, $\{ u, v \} \in E'$ and we conclude that
$\tE \subset E'$.  The theorem is proved.
\endpf

\begin{cor}
In Theorem~\r{renrvf}, if $V' = V - \{n\}$, then 
\[
E' = E^{\prime\prime} \cup \kappa(\phi(\{n\})) .
\]
\l{cor3}
\end{cor}

\pf
If suffice to observe that $\{n\}$ forms the only component of $G \bs V'$.
\endpf

\begin{eg}
Refer to Example~\r{eg8} and Fig.~\r{figeg8}.
Here $V-V' = \{ 3, 4, 7 \}$.  The components of $G \bs V'$ are
$\{ 3, 4 \}$ and $\{7\}$, and $\phi ( \{ 3, 4 \} ) = \{ 1, 2, 5, 6 \}$ and 
$\phi ( \{ 7 \} ) = \{ 2, 5, 8, 9 \}$.  Then 
\[
E' = E^\" \cup \kappa( \{ 1, 2, 5, 6 \} ) \cup \kappa(\{ 2, 5, 8, 9 \} ).
\]
\end{eg}

\begin{thm}
If $G \Rightarrow G' = (V', E')$, then $\{ u,v \} \in E'$ if 
there exists a path between $u$ and $v$ in $G$ on which all the intermediate vertices 
are in $V - V'$.
\l{a;lsrtj}
\end{thm}


\pf
Consider distinct $u,v \in V'$ such that 
there exists a path between $u$ and $v$ in $G$ on which all the intermediate vertices are in $V - V'$.
Denote this set of vertices in $V-V'$ by $S'$.  
Consider
\be
\tX_u \cap \tX_v - \tX_{V' - \{ u, v \}} 
=  \bigcup_{S \subset V - V'} \left( \tX_u \cap \tX_v  \cap \left( \bigcap_{t \in S} \tX_t \right) - \tX_{V - S - \{ u, v \} } \right) .
\l{qvnar}
\ee
Since $S' \subset V - V'$, we see that 
\[
A' = \tX_u \cap \tX_v  \cap \left( \bigcap_{t \in S'} \tX_t \right) - \tX_{V - S' - \{ u, v \}}
\] 
is one of the atoms in the union in (\r{qvnar}).  Note that $s(U_{A'}) = 1$ because $u$, $v$, and the vertices
in $S'$ form a path in $G$.  Thus $A'$ is a Type~I atom for $G$.

Now construct $X_i, i \in V$ by letting
\[
X_i = \left\{ \begin{array}{ll}
Z & \mbox{if $i \in S' \cup \{u, v\}$} \\
\mbox{constant} & \mbox{otherwise,}
\end{array} \right.
\]
where $Z$ is a random variable such that
$0 < H(Z) < \infty$.  Then by the proof of Theorem~3.11 in \cite {Yeung08},
for all $A \in \calA_V$,
\[
\mu^*(A) = \left\{ \begin{array}{ll}
H(Z) & \mbox{if $A = A'$} \\
0 & \mbox{otherwise.}
\end{array} \right.
\]
Now for $X_i, i \in V$ so constructed, $\mu^*$ vanishes on all the Type~II atoms of $G$
because $A'$, the only atom on which $\mu^*$ does not vanish, is a Type~I atom.  
Then from the discussion following Theorem~\r{qarocf}, we see that 
$X_i, i \in V$ satisfy~$[U_G]$.  On the other hand, in light of (\r{qvnar}), we have
\ba
\mu^* \left( \tX_u \cap \tX_v - \tX_{V' - \{ u, v \}} \right) 
& = & 
\sum_{S \subset V - V'} \mu^* \left( \tX_u \cap \tX_v  \cap \left( \bigcap_{t \in S} \tX_t \right) - \tX_{V -S - \{ u, v \}} \right) \\
& = & H(Z) \\
& > & 0 ,
\ea
i.e., $X_u$ and $X_v$ are not independent conditioning on $X_{V' - \{u,v\} }$.  
Hence, for any $G' = (V', E')$, if 
$G \Rightarrow G'$, then $V' - \{ u,v \}$ is not a cutset in $G'$, 
which implies that $\{ u, v\} \in E'$.
The theorem is proved.
\endpf

\bigskip
The next theorem is a rephrase of Theorem~\r{a;lsrtj} in light of the definition of $G^*(V')$ (Definition~\r{G*}).

\begin{thm}
If $G \Rightarrow G'$, then $G'$ contains $G^*(V')$ as a subgraph.
\l{wpujsv}
\end{thm}

\begin{thm}
$G \Rightarrow G^*(V')$.
\l{wpujsva}
\end{thm}

\pf
Let $X_i, i \in V$ be any set of random variables which satisfy $[U_G]$.
We need to prove that $X_i, i \in V'$ satisfy $\left[ U_{G^*(V')} \right]$.
For a fixed cutset $T \subset V'$ in $G^*(V')$, let $k$ be the number of components in $G^*(V') \bs T$
and denote these components by $Q_1, Q_2, \cdots, Q_k$.  To prove that $X_i, i \in V'$ 
satisfy $\left[ U_{G^*(V')} \right]$, 
it suffices to prove that for every cutset $T$ in $G^*(V')$,
$X_{Q_1}, X_{Q_2}, \cdots, X_{Q_k}$ are mutually independent conditioning on $X_T$.

Note that $\{ T, Q_1, Q_2, \cdots, Q_k \}$ is a partition of $\left( T \cup \left( \bigcup_i Q_i \right) \right) \subsetneq V$.
Following the discussion immediately after Theorem~\r{full_tatom}, we see that it suffices to prove that
$\mu^*$ vanishes on the sets prescribed in (\r{atom}).
The atoms of $\calF_V$ contained in a set prescribed in (\r{atom})
 have the form
\be
\left( \bigcap_{i=1}^k \bigcap_{j \in W_i} \tilde{X}_j \right) \cap  \left( \bigcap_{t \in S} \tX_t \right) - 
\tilde{X}_{T \cup ( \bigcup_{i=1}^k (Q_i - W_i) ) \cup (V-V'-S) } ,
\l{atom2}
\ee
where $S \subset V - V'$, $W_i \subset Q_i$, $1 \leq i \leq k$, and there exist at least two $i$ such that $W_i \neq \emptyset$. 

We will prove that every atom prescribed
in (\r{atom2}) is a Type~II atom of $G$.   Since $X_i, i \in V$ satisfy $[U_G]$, $\mu^*$
vanishes on these atoms.
It then follows that
\ba
\lefteqn{\mu^* \left( \left( \bigcap_{i=1}^k \bigcap_{j \in W_i} \tilde{X}_j \right) - 
\tilde{X}_{T \cup ( \bigcup_{i=1}^k (Q_i - W_i) ) }  \right) } \\
& = & 
\sum_{S \subset V-V'}
\mu^* \left(  \left( \bigcap_{i=1}^k \bigcap_{j \in W_i} \tilde{X}_j \right) \cap  \left( \bigcap_{t \in S} \tX_t \right) - 
\tilde{X}_{T \cup ( \bigcup_{i=1}^k (Q_i - W_i) ) \cup (V-V'-S)} \right) \\
& = & \sum_{S \subset V-V'} 0 \\
& = & 0 ,
\ea
i.e., $\mu^*$ vanishes on the sets prescribed in (\r{atom}), as is to be proved.

To prove that the atom in (\r{atom2}) is a Type~II atom of $G$, we need to show that 
$(T \cup ( \bigcup_{i=1}^k (Q_i - W_i) ) \cup (V-V'-S))$ is a cutset in $G$.  Now in (\r{atom2}),
let $1 \le i' < i^{\prime\prime} \le k$ be such that $W_{i'} \subset Q_{i'}$ and $W_{i^{\prime\prime}}
\subset Q_{i^{\prime\prime}}$ are nonempty, and let $u \in W_{i'}$ and $v \in W_{i^{\prime\prime}}$.
We claim that $u$ and $v$ are disconnected in $G \bs ( T \cup ( \bigcup_{i=1}^k (Q_i - W_i) ) \cup (V-V'-S) )$.  
Assume the contrary is true, i.e., there exists a path between $u$ and $v$ in $G \bs ( T \cup ( \bigcup_{i=1}^k (Q_i - W_i) ) \cup 
(V-V'-S) )$.
First of all, both $u$ and $v$ are in $V'-T$ and they belong to different components in $G^*(V') \bs T$.
Since $T \subset V' \subset V$, the vertices between $u$ and $v$ on this path are either in $V'-T$ or $V-V'$.
Then on this path (including $u$ and $v$) there exists two distinct vertices $w$ and $z$ in $V'-T$ such that 
\begin{list}%
{\arabic{cond})}{\usecounter{cond}}
\item
$w$ and $z$ are in different components in $G^*(V') \bs T$;
\item
all the vertices between $w$ and $z$ on the path
are in $V-V'$
\end{list}
(it is possible that $w = u$ and $z = v$).  Then 2) above implies that $\{ w, z \}$ is an edge in $G^*(V')$ (cf.\ Definition~\r{G*}),
which is a contradiction to 1).  Therefore, we conclude that $u$ and $v$ are disconnected in 
$G \bs ( T \cup ( \bigcup_{i=1}^k (Q_i - W_i) ) \cup (V-V'-S) )$.  Hence $G \bs ( T \cup ( \bigcup_{i=1}^k (Q_i - W_i) ) \cup (V-V'-S) )$
has at least two components and $(T \cup ( \bigcup_{i=1}^k (Q_i - W_i) ) \cup (V-V'-S))$ is a cutset in $G$.  
This completes the proof of the theorem.
\endpf

\bigskip
The following corollary gives a structural property of $G^*(V')$.

\begin{cor}
If $T$ is a cutset in $G^*(V')$, then $T$ is also a cutset in $G$.
\l{q5hcvv}
\end{cor}

\pf
In the proof of Theorem~\r{wpujsva}, we have proved that if $T$ is a cutset in $G^*(V')$, then 
$(T \cup ( \bigcup_{i=1}^k (Q_i - W_i) ) \cup (V-V'-S))$ is a cutset in $G$.
By setting $S = V-V'$ and $W_i = Q_i$ for all $i$, this cutset becomes $T$.
This proves the corollary.
\endpf

\bigskip
Combining Theorem~\r{wpujsv} and Theorem~\r{wpujsva}, we have proved the main result of this section.

\begin{thm}
Let $G= (V,E)$, and let $V' \subset V$.  Then $G^*(V')$ is the smallest $G'$ such that $G \Rightarrow G'$.
\l{quaff;ova}
\end{thm}


We end this section with a discussion.  There has been much research along the line of MRFs in the field of
graphical models \cite{Studeny05}.  In particular, classes of graphical models that 
contain undirected graph as a special case were defined in \cite{Sadeghi13, Sadeghi16},
where a separation 
criterion was provided for which the class of graphical models is stable under marginalization.  
In the context of the present paper, their result can be described as follows.  Let $G=(V,E)$ be an 
undirected graph and $V' \subset V$.
In \cite{Sadeghi13,Sadeghi16}, an algorithm is provided that takes $G$ as the input and produces a graph 
as the output which is essentially the same as $G^*(V')$, and it was shown that
if $X_i, i \in V$ satisfy only those conditional independencies
induced by $G$ (i.e., $X_i, i \in V$ satisfy the conditional independencies
induced by $G$ and no more), then $X_i, i \in V'$ satisfy only those conditional independencies induced by $G^*(V')$.
This implies that if $G \Rightarrow G'$, then $G'$ cannot be a subgraph of $G^*(V')$, i.e., Theorem~\r{quaff;ova}.

Although the graph produced by the algorithm in \cite{Sadeghi16} is essentially the same as
$G^*(V')$, it is not given in closed form.
By contrast, our closed-form characterizations of $G^*(V')$ (Definition~\r{G*}, Theorem~\r{renrvf},
and Corollary~\r{cor3}) facilitate the development of further results, including
Proposition~\r{V"} and the recursive approach for constructing information diagrams for MRFs to be discussed in 
Section~\r{sec6}.

It is also worth pointing out that our proof of Theorem~\r{quaff;ova}, which is information-theoretic,
is interesting on its own because it is developed upon the view that an MRF is a collection of FCMIs.
As such, some of the results in this paper can potentially be generalized for general collections
of FCMIs, which is beyond the scope of graphical models.
 
\section{Markov Tree}
\l{sec4}
Suppose $X_i, i \in V$ are represented by a graph $G$.  If $G$ is a tree, then $X_i, i \in V$ form a Markov tree.  
If $G^*(V')$ is also a tree, we say that $X_i, i \in V'$ form a Markov subtree.  For the special case when $G$ is a path,
it is easy to see that $G^*(V')$ is always a path (see Example~\r{eg7} for instance).  In other words, if $X_i, i \in V$ form a Markov chain, then for any 
$V' \subset V$, $X_i, i \in V'$  always form a Markov subchain.

However, if $X_i, i \in V$ form a Markov tree, for an arbitrary subset $V'$ of $V$, $X_i, i \in V'$ may or may not form a Markov subtree.
The following theorem, which is an application of Theorem~\r{quaff;ova}, gives a necessary and sufficient condition for $X_i, i \in V'$
to form a Markov subtree. 

\begin{thm}
Let $X_i, i \in V$ form an MRF represented by a tree $G = (V,E)$.  
For $V' \subset V$, $G^*(V')$ is a tree if and only if there do not exist
$u \in V-V'$ and $v_1, v_2, v_3 \in V'$ such that for $i = 1, 2, 3$, all the vertices on the path between $u$ and $v_i$
except for $v_i$ are in $V - V'$.  
\l{vntierqe}
\end{thm}

\pf
We first prove the ``only if" part.  Assume that $G^*(V') = (V', E')$ is a tree and there exist
$u \in V-V'$ and $v_1, v_2, v_3 \in V'$ and such that for $i = 1, 2, 3$, all the vertices on the path between $u$ and $v_i$
except for $v_i$ are in $V - V'$.  By Definition~\r{G*}, the edges $(v_1, v_2)$, $(v_2, v_3)$, and $(v_1, v_3)$
are in $E'$.  Hence $v_1, v_2, v_3$ form a cycle in $G^*(V')$, a contradiction to the assumption that
$G^*(V')$ is a tree.

We now prove the ``if" part.  Assume that $G^*(V') = (V', E')$ is not a tree.
Then there exists a cycle $w_0, w_1, \cdots, w_{m-1}, w_0$ in $G^*(V')$, where $m \ge 3$ and 
  $w_0, w_1, \cdots, w_{m-1} \in V'$ are distinct.  By Definition~\r{G*}, for each $0 \le i \le m-1$, there exists
a path between $w_i$ and $w_{i+1}$ in $T$ on which all the intermediate vertices are in $V - V'$, 
where `+' in the subscript denotes modulo~$m$ addition.  This path is in fact
unique because $G$ is a tree, so we denote it by \Path$(w_i, w_{i+1})$.

If all the vertices on the collection of paths \Path$(w_i, w_{i+1})$, $0 \le i \le m-1$, except for the endpoints, are distinct, 
since $w_0, w_1, \cdots, w_{m-1}$ are distinct, these paths together form a cycle in $T$ which is a contradiction
because $T$ is a tree.  
Otherwise, there exists a vertex 
$u \in V-V'$ which is on both
\Path$(w_i, w_{i+1})$ and \Path$(w_j, w_{j+1})$ for some $0 \le i < j \le m-1$.
Note that $| \{ w_i, w_{i+1} \} \cup \{ w_j, w_{j+1} \} | \ge 3$, with equality if and only if $j = i+1 \mod m$.
Then there exist $v_1, v_2, v_3 \in \{ w_i, w_{i+1} \} \cup \{ w_j, w_{j+1} \} \subset V'$
such that for $i = 1, 2, 3$, all the vertices on the path between $u$ and $v_i$
except for $v_i$ are in $V - V'$.  The theorem is proved.
\endpf

\begin{eg}
Consider a Markov tree represented by the tree $G$ in Fig.~\r{figeg10} 
and let $V' = \{ 1, 4, 8, 9, 12 \}$.  The graph $G^*(V')$, illustrated as the overlay graph
in grey, is evidently a tree.  We call $G^*(V')$ a Markov subtree.
It can be checked that the condition in Theorem~\r{vntierqe} is satisfied.\

However, if $V'$ also includes vertex~7, then $G^*(V')$ as shown in Fig.~\r{figeg10b} is not a tree.
By letting 
$u = 6$, $v_1 = 4$, $v_2 = 7$, and $v_3 = 8$, we see that the condition
in Theorem~\r{vntierqe} is violated because $u$ is connected to each of $v_1$, $v_2$, and $v_3$ by an edge in 
$V - V'$.
\l{eg10}
\end{eg}

\section{Markov Chain}
\l{sec5}
A Markov chain is a special case of a Markov tree.
However, there are certain properties that are possessed by a Markov chain but not by a Markov tree in general.
Consider the graph $P_n = (V, E)$, where $V = \{ 1, 2, \cdots, n \}$ and the edges in $E$ are
$\{ i, i+1 \}$ for $ i = 1, 2, \cdots, n-1$.  Evidently, $P_n$ is a path.  If $X_i, i \in V$ is represented by $P_n$, then $X_i, i \in V$
form the Markov chain $X_1 \ra X_2 \ra \cdots \ra X_n$.
The following properties of a (finite-length) Markov chain were proved in \cite{KawabataY92}:
\begin{list}%
{(C\arabic{cond})}{\usecounter{cond}}
\item
An atom $A$ of $\calF_V$ is a Type~I atom if and only if 
\be
U_A = V - \{ l, l+1, \cdots, u \}
\l{vnaern}
\ee
where $1 \le l \le u \le n$, i.e., the indices of the set variables in $A$ that are not complemented are consecutive.
\item
The values of $\mu^*$ on all the Type~I atoms are nonnegative.
\item
$\mu^*$ vanishes on all the Type~II atoms.
\end{list}
Since $\mu^*$ vanishes on all the Type~II atoms and is nonnegative on all the Type~I atoms,
it is a measure on $\calF_V$.  Also, the $I$-Measure $\mu^*$ of a finite-length
Markov chain can be represented by a 2-dimensional information diagram as in Fig.~\r{figMC}, in which 
all the Type~II atoms are suppressed.
 
Subsequently, (C3) was generalized for arbitrary finite undirected graphs \cite{YeungLY02}.  However, the nonnegativity of 
the $I$-Measure does not hold even for the simplest Markov tree that is not 
a Markov chain \cite{YeungLY02}.

\begin{eg}
Let $Z_1$ and $Z_2$ be i.i.d.\ random variables each distributed uniformly
on $\{0,1\}$.  Let $X_1 = Z_1$, $X_2 = Z_2$,
$X_3 = Z_1 + Z_2 \mod 2$, and $X_4 = (Z_1,Z_2)$.  Since $X_1$, $X_2$, and $X_3$ 
are functions of $X_4$,
they are mutually independent conditioning on $X_4$.
Thus $X_1, X_2, X_3$, and $X_4$ form a Markov tree represented by the ``star"
in Figure~\r{figeg11}.  It is not difficult to show that (see \cite[Example~3.10]{Yeung08})
\[
\mu^*(\tX_1 \cap \tX_2 \cap \tX_3) = -1,
\]
and hence $\mu^*$ is not nonnegative.
\l{eg11}
\end{eg}

Before explaining the significance of the nonnegativity of $\mu^*$ for Markov chains,
we first review the following result in \cite{KawabataY92} which is instrumental in proving the 
nonnegativity of $\mu^*$ for a Markov chain.
Prior to \cite{KawabataY92}, the same result (and also the converse) was proved in \cite{Hu62} for the special case
$U_A = \emptyset$.  

\begin{lemma}
If $X_1 \ra X_2 \ra \cdots \ra X_n$ form a Markov chain, then for a Type~I atom with $U_A$ defined in (\r{vnaern}),
\be
\mu^*(A) = \mu^* \left( \tX_l \cap \tX_{l+1} \cap \cdots \cap \tX_u - \tX_{U_A} \right) = 
\mu^* \left(  \tX_l \cap \tX_u - \tX_{U_A} \right) .
\l{a;5hvoa}
\ee
\l{lem:viaorh}
\end{lemma}

Note that the first equality above follows directly from the definition of $U_A$, and the quantity
on the right hand side is 
equal to $I( X_l ; X_u | X_{U_A} )$ which is always nonnegative.  In other words, Lemma~\r{lem:viaorh} asserts that the values of
$\mu^*$ on all the Type~I atoms are nonnegative.  Therefore, $\mu^*$ is a measure.

As mentioned in Section~\r{sec2.1}, for all $A \in \calA_V$, $\mu^*(A)$ is a linear combination of $H(X_B)$ for nonempty subsets $B$ of $V$.  Then if $X_1 \ra X_2 \ra \cdots \ra X_n$ forms a Markov chain,
any linear information inequality involving $X_i, i \in V$
can be expressed in the form
\[
\sum_{A \, \in \, {\calT}_{\rm I}} c_A \, \mu^*(A) \ge 0 ,
\]
where $c_A \in \mathbb{R}$.  The following theorem gives a complete characterization of
such inequalities that always holds.

\begin{thm}
If $X_1 \ra X_2 \ra \cdots \ra X_n$ forms a Markov chain, then
\be
\sum_{A \, \in \, {\calT}_{\rm I}} c_A \, \mu^*(A) \ge 0 
\l{atcoiai}
\ee
always holds if and only if $c_A \ge 0$ for all $A \in {\calT}_{\rm I}$.
\l{Gamma}
\end{thm}
 
\pf
If $X_1 \ra X_2 \ra \cdots \ra X_n$ forms a Markov chain, then $\mu^*(A) \ge 0$ for all
$A \, \in \, {\calT}_{\rm I}$.  If $c_A \ge 0$ for all $A \in {\calT}_{\rm I}$, then evidently
(\r{atcoiai}) always holds.

To prove the converse, assume that $c_{A'} < 0$ for an atom $A' \in {\calT}_{\rm I}$.
Now construct $X_i, i \in V$ by letting
\[
X_i = \left\{ \begin{array}{ll}
Z & \mbox{if $i \in U_{A'}$} \\
\mbox{constant} & \mbox{otherwise,}
\end{array} \right.
\]
where $Z$ is a random variable such that
$0 < H(Z) < \infty$.  Then by the proof of Theorem~3.11 in \cite {Yeung08},
for all $A \in \calT_{\rm I}$,
\[
\mu^*(A) = \left\{ \begin{array}{ll}
H(Z) & \mbox{if $A = A'$} \\
0 & \mbox{otherwise.}
\end{array} \right.
\]
It follows that
\[
\sum_{A \, \in \, {\calT}_{\rm I}} c_A \, \mu^*(A) = c_{A'} \, \mu^*(A') < 0 
\]
since $c_{A'} < 0$ and $\mu^*(A') > 0$.  Hence, (\r{atcoiai}) does not always hold and 
the converse is proved.
\endpf 

\bigskip
\noindent
{\bf Remark} \
Let $X_1 \ra X_2 \ra \cdots \ra X_n$ form a Markov chain and consider any inequality of the form ({\r{atcoiai}) that
always holds.  Theorem~\r{Gamma} asserts that the left hand side of ({\r{atcoiai}) must be a conic
combination of $\mu^*(A), A \in \TI$.  Since $\mu^*(A)$ is a Shannon's information 
measure for all $A \in \TI$, we see that ({\r{atcoiai}) is implied by the nonnegativity of Shannon's
information measures and hence is a {\em Shannon-type information 
inequality} (see \cite[Ch.~14]{Yeung08}).  Therefore, we conclude that there exist no {\em non-Shannon-type} 
information inequalities for a Markov chain.

\bigskip
Fix a graph $G$ and let ${\cal P}_G$ be the 
class of probability measures $P$ on $\calX_1 \times \calX_2 \times \cdots \times \calX_n$
such that $P$ forms an MRF represented by $G$.
In the rest of this section, we prove that  
the $I$-Measure $\mu^*$ of every $P \in {\cal P}_G$ is nonnegative if and only
if $G$ is either a path or a forest of paths.
In other words, the MRF represented by such a graph $G$ is either a Markov chain or 
a collection of mutually independent Markov chains.
In this sense we say that the Markov chain is the only MRF for which the $I$-Measure is always nonnegative.

In the following, we present a theorem which is a generalization of Lemma~\r{lem:viaorh}.  
Unlike Lemma~\r{lem:viaorh} that applies only to Markov chains, this theorem applies to all MRFs.

\begin{thm}
Let $X_1, X_2, \cdots, X_n$ form a Markov graph $G = (V,E)$.
For a Type~I atom $A$ of $G$ with $|U_A| \le n-2$,  
\be
\mu^*(A) 
= \mu^* \left( \bigcap_{k \in B} \tX_k - \tX_{U_A}  \right) ,
\l{2n09jf}
\ee
where 
\[
B = \{ k \in V - U_A : s( U_A \cup \{ k \} ) = 1 \} ,
\]
i.e., a vertex $k \not\in U_A$ is in $B$ if and only if upon
removing all the vertices in $U_A$ and vertex $k$, the graph remains connected.
\l{kfklaelr}
\end{thm}

\begin{eg}
Consider an MRF represented by the graph in Fig.~\r{figeg12}.
For the Type~I atom $\bar{1}2\bar{3}45678$, using Theorem~\r{kfklaelr}, 
$B = \{ 2, 7, 8 \}$, and so
\[
\mu^*\left( \tX_2 \cap \tX_4 \cap \tX_5 \cap \tX_6 \cap \tX_7 \cap \tX_8 - \tX_{\{1,3\}} \right)
= \mu^*\left( \tX_2 \cap \tX_7 \cap \tX_8 - \tX_{\{1,3\}} \right).
\]
For the Type~I atom $\bar{1}\bar{2}345467\bar{8}$, $B = \{ 3, 4, 6, 7 \}$, and so
\[
\mu^*\left( \tX_3 \cap \tX_4 \cap \tX_5 \cap \tX_6 \cap \tX_7 - \tX_{\{1,2,8\}} \right)
= \mu^*\left( \tX_3 \cap \tX_4 \cap \tX_6 \cap \tX_7 - \tX_{\{1,2,8\}} \right).
\]

\l{eg12}
\end{eg}

To gain insight into Theorem~\r{kfklaelr}, we first state the next lemma.
This lemma and the technical lemma that follows will be proved in Appendix~\r{App_B}.

\begin{lemma}
In Theorem~\r{kfklaelr}, $|B| \ge 2$.
\l{vanetl}
\end{lemma}

\noindent
{\bf Remark} \ When $|B| = 2$, the term on the right hand side of (\r{2n09jf}) becomes a (conditional) mutual information,
which is always nonnegative.

\bigskip
The following lemma will be used in the proof of Theorem~\r{kfklaelr}.
 
\begin{lemma}
In Theorem~\r{kfklaelr}, let $W = V-U_A-B$.  For any $S \subsetneq W$,
$s(U_A \cup (W-S)) > 1$.
\l{kffgjlj}
\end{lemma}

\medskip
\noindent
{\bf Proof of Theorem~\r{kfklaelr}} \
Let $W = V - U_A - B$.  Consider
\ba
\mu^* \left( \bigcap_{k \in B} \tX_k - \tX_{U_A}  \right)
& = & \mu^* \left( \bigcup_{S \subset W} \left( 
\left( 
\bigcap_{k \in B} \tX_k 
\right) 
\cap 
\left( 
\bigcap_{t \in S} \tX_t 
\right) 
- \tX_{U_A \cup (W - S)} \right)  \right)\\
& = & \sum_{S \subset W} \mu^* \left( \left( \bigcap_{k \in B} \tX_k \right) \cap 
\left( \bigcap_{t \in S} \tX_t \right) - \tX_{U_A \cup (W - S)} \right).
\ea
In the above summation, for $S \subsetneq W$, $s(U_A \cup (W - S)) > 1$
by Lemma~\r{kffgjlj}.  Therefore, except for the atom corresponding to $S = W$, i.e., $A$,
all the atoms are Type~II atoms of $G$.  
It then follows that 
\ba
\mu^* \left( \bigcap_{k \in B} \tX_k - \tX_{U_A}  \right)
& = & \mu^* \left( \left( \bigcap_{k \in B} \tX_k \right) \cap 
\left( \bigcap_{t \in W} \tX_t \right) - \tX_{U_A} \right) \\
& = &  \mu^* \left(  \left( \bigcap_{k \in V-U_A} \tX_k \right) - \tX_{U_A} \right) \\
& = & \mu^*(A) .
\ea
The theorem is proved.
\endpf

\bigskip
Theorem~\r{kfklaelr} can be applied to identify atoms on which the value of $\mu^*$ is
always nonnegative, because when $|B|=2$, the term on the right hand side of (\r{2n09jf})
corresponds
to a (conditional) mutual information.

Consider the graph $G = (V,E)$, where $V = \{ 1, 2, \cdots, n \}$ 
and the edges in $E$ are $\{ i, i+1 \}$ for $ i = 1, 2, \cdots, n-1$ and $\{ 1, n \}$.  Evidently,
$G$ is a {\em cycle graph}, and if random variables $X_i, i \in V$ are represented by $G$, they
form a {\em Markov ring}.  Then $A$ is a Type~I atom of $G$ if and only if $U_A = \emptyset$ or $U_A$ is a consecutive 
subset of $V$ in the cyclic sense (e.g., $\{ 1, n \}$ is regarded as a consecutive subset of $V$).
An application of Theorem~\r{kfklaelr} reveals that 
$\tX_1 \cap \tX_2 \cap \cdots \cap \tX_n$ (i.e., $U_A = \emptyset$)
is the only atom on which $\mu^*$ may 
take a negative value, because the value of $\mu^*$ on any other Type~I atom is a conditional mutual information.
This observation is instrumental in the proof of the next theorem, the main result in this section.


\begin{thm}
Let $G$ be a connected graph. Then $\mu^*$ is nonnegative for every $P \in {\cal P}_G$ if and only if
$G$ is a path.
\l{aiog;o}
\end{thm}

The `if' part of Theorem~\r{aiog;o} is immediate because the $I$-Measure 
for a Markov chain is always nonnegative.
Toward proving the `only if' part, we first classify all connected graphs into the following two classes:
\begin{description}
\item
\hspace{.25in} $K1$: there exists a vertex whose degree is at least 3;
\item
\hspace{.25in} $K2$: all the vertices have degree less than or equal to 2.
\end{description}
We further classify the graphs in $K2$ into two subclasses:
\begin{description}
\item
\hspace{.25in} $K2$-$a$: all the vertices have degree 2;
\item
\hspace{.25in} $K2$-$b$: some vertices have degree 1.
\end{description}

It is easy to see that a graph belonging to subclass $K2$-$a$ is a cycle graph, and a graph belonging
to subclass $K2$-$b$ is a path.  Thus in order to establish Theorem~\r{aiog;o}, it suffices to prove Theorem~\r{thm9} and Theorem~\r{thm10} below which assert
that $\mu^*$ is not always nonnegative if 
$X_i, i \in V$ are represented by a graph belonging to $K1$ and $K2$-$a$, respectively.

\begin{thm}
The $I$-Measure $\mu^*$ for an MRF represented by a graph $G$ belonging to $K1$ is not always nonnegative.
\l{thm9}
\end{thm}

\pf
Consider a graph $G = (V,E)$ in $K1$.  Let $u \in V$ be a vertex whose degree is at least 3, and let
$\{ u,v_l \} \in E$, where $l = 1, 2, 3$ and $v_1, v_2$, and $v_3$ are distinct.  Let $Z$ and $T$ be independent fair bits.  Define random variables
$X_i, i \in V$ as follows:
\[
X_i = \left\{ \begin{array}{ll}
Z & i = v_1 \\
T & i = v_2 \\
Z + T \ \mbox{mod} \ 2 & i = v_3 \\
(Z,T) & i = u \\
\mbox{constant} & \mbox{otherwise}
\end{array} \right.
\]
Consider any cutset $U$ of $G$:
\begin{enumerate}
\item
If $u \not\in U$, then $u$ and $v$ for all $v \in \{ v_1, v_2, v_3 \} \bs U$ are in the same component of $G \bs U$ because
$u$ and $v$ are connected by an edge in $G$.
Since $X_i = \mbox{constant}$ for all $i \ne u, v_1, v_2, v_3$,
it is readily seen that $X_{V_1(U)}, X_{V_2(U)}, \cdots, X_{V_{s(U)}(U)}$ are 
mutually independent conditioning on $X_U$.
\item
If $u \in U$, since $X_{v_l}, l = 1, 2, 3$ are functions of $X_u$ and 
$X_i = \mbox{constant}$ for all $i \ne u, v_1, v_2, v_3$, it is readily seen that 
$X_{V_1(U)}, X_{V_2(U)}, \cdots, X_{V_{s(U)}(U)}$ are 
mutually independent conditioning on $X_U$.
\end{enumerate}
Thus in either case $X_i, i \in V$ are represented by $G$.  Then
\ba
\mu^* \left( \tX_u \cap \tX_{v_1} \cap \tX_{v_2} \cap \tX_{v_3} - \bigcup_{i \ne  u, v_1, v_2, v_3} \tX_i  \right) 
& = & \mu^*(\tX_u \cap \tX_{v_1} \cap \tX_{v_2} \cap \tX_{v_3}) \\
& = & -1 ,
\ea
where the first equality can be seen by expanding 
$\mu^* 
\left( \tX_u \cap \tX_{v_1} \cap \tX_{v_2} \cap \tX_{v_3} - \bigcup_{i \ne  u, v_1, v_2, v_3} \tX_i  \right)$ 
using \cite[Theorem~3.19]{Yeung08} into 
a linear combination of $H ( \, \cdot \, | \, \tX_i, i \ne  u, v_1, v_2, v_3) = H( \, \cdot \, )$, and 
the second equality can easily be verified (cf.\ Problem~5, Ch.~12 in \cite{Yeung08}).  Hence, $\mu^*$ for $X_i, i \in V$
represented by a graph $G$ belonging to $K1$ is not always nonnegative.

\begin{thm}
The $I$-Measure $\mu^*$ for an MRF represented by a graph $G$ belonging to $K2$-$a$ is not always nonnegative.
\l{thm10}
\end{thm}
 
\pf
Consider a graph $G = (V,E)$ in $K2$-$a$, i.e., $G$ is a cycle graph.
For convenience, let $V = \{ 0, 1, \cdots, n-1 \}$.  The edge set $E$ is specified by
$\{ u,v \} \in E$ if and only if $|u-v| = 1$, where ``$-$" denotes modulo~$n$ subtraction.
Let $F$ denote a finite field containing at least $n-1$ elements.  Let $Z$ and $T$ be independent
random variable, each taking values in $F$ according to the uniform distribution.
Now define random variables $X_i, i \in V$ as follows:
\[
X_i = \left\{ \begin{array}{ll}
Z & i = 0 \\
T & i = 1 \\
Z + \alpha_i T  & i = 2, 3, \cdots, n-1
\end{array} \right.
\]
where $\alpha_i, i = 2, 3, \cdots, n-1$ are distinct nonzero elements of $F$.  It is evident that
$X_i, i = 0, 1, \cdots, n-1$ are pairwise independent but not mutually independent, and 
that for any distinct $i, i', i^\"$, we have $X_{i^\"}$ being a function of $(X_i, X_{i'})$.

We now show that $X_i, i \in V$ is represented by $G$.  Since $G$ is a cycle graph, for any $U \subset V$,
if the vertices in $U$ are connected in $G$, the vertices in $V - U$ are also connected in $G$.  Therefore, if
$U$ is a cutest in $G$, the vertices in $U$ are not connected in $G$.  This implies that $|U| \ge 2$.
From the foregoing, $X_{V - U}$ is a function of $X_U$.
Then we see that $X_{V_1(U)}, X_{V_2(U)}, \cdots, X_{V_{s(U)}(U)}$ are 
mutually independent conditioning on $X_U$.  Therefore, $X_i, i \in V$ is represented by $G$.

It remains to show that $\mu^*$ is not nonnegative.  
For the sake of convenience, assume the logarithms defining entropy are in the base $|F|$.
Then for $B \subset V$ such that $B \ne \emptyset$,
\be
H(X_B) = \left\{ \begin{array}{ll}
1 & \mbox{if $|B| = 1$} \\
2 & \mbox{if $2 \le |B| \le n$}.
\end{array} \right.
\l{q89uy5}
\ee
We will show that 
$\mu^*$ is given by
\be
\mu^* \left( \bigcap_{i \in W} \tX_i - \bigcup_{j \in V-W} \tX_j \right) = \left\{ \begin{array}{ll}
0 & \mbox{if $1 \le |W| \le n-2$} \\
1 & \mbox{if  $|W| = n-1$} \\
-(n-2) & \mbox{if $|W| = n$}
\l{98hn;oat}
\end{array} \right.
\ee
for $W \subset V$.
Toward this end, owing to the uniqueness of $\mu^*$, we only need to verify that
$\mu^*$ as prescribed by (\r{98hn;oat}) satisfies (\r{q89uy5}).
The details are given in Appendix~\r{App_A}.
Then the theorem is proved because $\mu^*$ is not nonnegative.
\endpf

\begin{thm}
Let $G$ be a graph with at least two components.  Then $\mu^*$ is nonnegative
for every $P \in {\cal P}_G$ if and only if $G$ is a forest of paths. 
\end{thm}

\pf
We first prove the `only if' part.  Assume that $G$ is not a forest of paths, i.e., 
there exists a component of $G$ which is not a path.  Denote the vertices 
of this component by $V'$ and let $X_i, i \in V \bs V'$ be constant.  Then by Theorem~\r{aiog;o}, we can construct $X_i, i \in V$ such that
$\mu^*(S) < 0$ for some $S \subset \calF_{V'} \subset \calF_V$,
where $\calF_{V'}$ is the $\sigma$-field generated by $\{ \tX_i, i \in V' \}$.
Hence $\mu^*$ is not nonnegative, and the `only if' part is proved.

To prove the `if' part, we need to prove that if an MRF is represented by a graph $G$
which is 
a forest of paths, then $\mu^*$ is always nonnegative.
Let $m$ be the number of components of $G$, where $m \ge 2$, and
denote the sets of vertices of these components by 
$V_1, V_2, \cdots, V_m$.  Without loss of generality, assume that
the indices in each $V_i$ are consecutive.

Now observe that a nonempty atom $A$ of $\calF_V$ is a Type~I atom
of $\calF_V$ if and only if $U_A$ has the form (\r{vnaern}) and
$\{ l, l+1, \cdots, u \} \subset V_i$ for some $1 \le i \le m$.
If $l = u$, then 
\[
\mu^*(A) = H(\tX_l | \tX_{V - \{l\}}) \ge 0.
\]
If $l < u$, then by Theorem~\r{kfklaelr}, 
\[
\mu^*(A) = \mu^* \left(
 \bigcap_{l \le k \le u} \tX_k - \tX_{U_A} \right)
= \mu^*(\tX_l \cap \tX_k - \tX_{U_A} ) \ge 0.
\]
Hence $\mu^*$ is nonnegative, and the theorem is proved.
\endpf

\section{Information Diagrams for Markov Random Fields}
\l{sec6}
As discussed in Section~\r{sec5}, the $I$-Measure $\mu^*$ of a finite-length
Markov chain can be represented by a 2-dimensional information diagram as in Fig.~\r{figMC}.  
Such an information diagram is a ``correct" representation
in the sense that the closed curves representing the set variables
intersect with each other in such a way that
\begin{enumerate}
\item
all the Type~I atoms are nonempty (not suppressed);
\item
all the Type~II atoms are empty (suppressed).
\end{enumerate}
We call Fig.~\r{figMC} an information diagram (customized) for a Markov chain,
or more specifically an information diagram for the path $P_n$ (as discussed in Section~\r{sec5}).
With such an information diagram, it is relatively easy to discover 
information inequalities and identities 
pertaining to a Markov chain by visualization, which may be difficult otherwise.
A notable such example is an information identity for a Markov chain of five random variables
that was useful in proving an outer bound for multiple descriptions \cite{FuY02} (see also
\cite[Example~3.18]{Yeung08}).

Owing to its simple and regular structure, it is possible to construct an
information diagram for a Markov chain by trial and error.
However, constructing an information diagram for a general MRF requires a more systematic approach.
In the rest of this section, we develop a method for this purpose by using the characterization of a subfield of an MRF in Section~\r{sec4}. 

To simplify notation, we use $N_n$ to denote $\{1, 2, \cdots, n\}$.  
Consider $X_i, i \in V$ forming a Markov graph
$G = (V, E)$ with $V = N_n$.  Using Corollary~\r{cor3} as the recipe, we can construct
$G^*(N_{n-1})$.  Then by repeating this step with $G^*(N_{n-1})$ in place of $G$,
we can construct $(G^*(N_{n-1}))^*(N_{n-2})$, which from Proposition~\r{V"} is in fact equal to $G^*(N_{n-2})$.
In the same fashion, we can construct the graphs $G^*(N_{n-3}), \cdots, G^*(N_{1})$ recursively.

In our method for constructing an information diagram for
$G$, we construct a sequence of information diagrams for $G^*(N_1)$,
$G^*(N_{2}), \cdots, G^*(N_n) = G$ recursively, with the last one being the desired information diagram.
Denote these information diagrams by $\calD_1, \calD_2, \cdots, \calD_n$.
For the convenience of discussion, denote the closed curve representing $\tX_m$ by $\calC_m$ for 
$1 \le m \le n$.

Now the graph $G^*(N_{1})$ consists of the single vertex 1 and no edge.
Then an information diagram consisting of any closed curve representing $\tX_1$ would be 
a correct representation for $G^*(N_{1})$.  Call this information diagram $\calD_1$.

We observe that for $m = 2, 3, \cdots, n$, an atom $A \in \calA_{N_{m-1}}$ generates
the two atoms $A \cap \tX_{m}$ and $A \cap \tX_{m}^c$ in $\calA_{N_{m}}$, and there is the 
extra atom $\tX_1^c \cap \tX_2^c \cap \cdots \cap \tX_{m-1}^c \cap \tX_{m}$ in $\calA_{N_{m}}$ 
that is not generated by any atom in $\calA_{N_{m-1}}$.

For $m = 2, 3, \cdots, n$, in constructing $\calD_{m}$ from $\calD_{m-1}$, we add the closed curve  
$\calC_{m}$ to the former in a suitable way.
In order for this recursive approach to work, we need to ensure that a Type~II atom of $G^*(N_{m-1})$
that is suppressed in $\calD_{m-1}$ would not generate a Type~I atom of $G^*(N_{m})$ which is not to be suppressed
in $\calD_{m}$.  This is proved in the next theorem.

\begin{thm}
For $m = 2, 3, \cdots, n$, if $A$ is a Type~II atom of $G^*(N_{m-1})$, then both $A \cap \tX_{m}$ and $A \cap \tX_{m}^c$ are 
Type~II atoms of $G^*(N_{m})$.
\l{GNm}
\end{thm}

\pf
Assume that $A$ is a Type~II atom of $G^*(N_{m-1})$. We first prove that $A \cap \tX_{m}^c$ is a Type~II atom of $G^*(N_{m})$. 
From the discussion following Proposition~\r{V"}, we know that $G^*(N_{m}) \bs \{m\}$ is a subgraph of
$G^*(N_{m-1})$.  As such, upon removing all the vertices in $U_A$ in both graphs, 
we see that $G^*(N_{m}) \bs (\{m\} \cup U_A)$ is subgraph of $G^*(N_{m-1}) \bs U_A$, where the latter is disconnected
because $A$ is a Type~II atom of $G^*(N_{m-1})$.
It then follows that $G^*(N_{m}) \bs (\{m\} \cup U_A)$ is also disconnected.  
Upon noting that $U_{A \, \cap \, \tX_m^c} = U_A \cup \{m\}$, we see that
$G^*(N_{m}) \bs U_{A \cap \tX_m^c} = G^*(N_{m}) \bs (\{m\} \cup U_A)$ which is disconnected.
Therefore, 
$A \cap \tX_{m}^c$ is a Type~II atom of $G^*(N_{m})$.

We now prove that $A \cap \tX_{m}$ is a Type~II atom of $G^*(N_{m})$. 	
Let
\be
\gamma_m = \left\{ \, j : \mbox{$\{j, m\}$ is an edge in $G^*(N_{m})$} \, \right\}
\l{gammam}
\ee
be the set of neighbors of vertex $m$ in $G^*(N_{m})$.  Note that $\gamma_m \subset N_{m-1}$.
Let $\tE_m$ and $\tE_{m-1}$ be the sets of edges of $G^*(N_{m}) \bs U_A$ and $G^*(N_{m-1}) \bs U_A$,
respectively.  Evidently,
\be
\tE_m - \tE_{m-1} = \left\{ \{m,j\} : j \in \gamma_m - U_A \right\} ,
\l{gerfvss}
\ee
where $\gamma_m -U_A$ is the set of neighbors of vertex
$m$ in $G^*(N_m) \bs U_A$.
We consider two cases for $\gamma_m -U_A$.

\noindent
\underline{$\gamma_m -U_A = \emptyset$}

\noindent
This is the case when vertex $m$ has no neighbor in $G^*(N_m) \bs U_A$.
Since $A$ is a Type~II atom of $G^*(N_{m-1})$, $G^*(N_{m-1}) \bs U_A$ is disconnected.
From (\r{gerfvss}), we have $ \tE_m - \tE_{m-1} = \emptyset$, so that $\tE_m \subset \tE_{m-1}$.
This implies that $G^*(N_{m}) \bs U_A$ is also disconnected.
Upon noting that $U_{A \cap \tX_m} = U_A$, we have $G^*(N_{m}) \bs U_{A \cap \tX_m} = G^*(N_{m}) \bs U_A $
which is disconnected.  Therefore, $A \cap \tX_{m}$ is a Type~II atom of $G^*(N_{m})$.

\noindent
\underline{$\gamma_m -U_A \ne \emptyset$}

\noindent
This is the case when vertex $m$ has at least one neighbor in $G^*(N_m) \bs U_A$.
For any  distinct vertices $i_1$ and $i_2$ that are both neighbors of vertex~$m$ 
in $G^*(N_m) \bs U_A$ (and therefore also neighbors of vertex~$m$ 
in $G^*(N_m)$), according to Corollary~\r{cor3}, 
$\{i_1, i_2\}$ is an edge in $G^*(N_{m-1})$ and hence $\{i_1, i_2\} \in \tE_{m-1}$ (because
$i_1, i_2 \not\in U_A$), which implies that $i_1$ and $i_2$ belong to 
the same component in $G^*(N_{m-1}) \bs U_A$.  
Equivalently, if $i_1$ and $i_2$ belong to different components in $G^*(N_{m-1}) \bs U_A$,
then $i_1$ and $i_2$ cannot both be neighbors of $m$ in $G^*(N_m) \bs U_A$.
Therefore, the neighbors of
$m$ in $G^*(N_m) \bs U_A$ all belong to the same component in $G^*(N_{m-1}) \bs U_A$,
and we denote this component by $V_1$.

Since $G^*(N_{m-1}) \bs U_A$ is disconnected, there exists another component $V_2$ in $G^*(N_{m-1}) \bs U_A$.
Now consider any $i_1 \in V_1$ and $i_2 \in V_2$.  Since $i_1$ and $i_2$ are in different components in
$G^*(N_{m-1}) \bs U_A$, we have $\{i_1, i_2\} \not\in \tE_{m-1}$.  Then
we see from the discussion in the last paragraph that $i_1$ and $i_2$ cannot both be 
neighbors of $m$ in $G^*(N_m) \bs U_A$.   
By Corollary~\r{cor3}, $\{i_1, i_2\}$ is not an edge in $G^*(N_{m-1})$ and hence not an edge in 
$G^*(N_m) \bs U_A$ because $G^*(N_m) \bs U_A$ is a subgraph of  $G^*(N_{m-1})$.
Therefore,
$(i_1, i_2) \not\in \tE_{m}$.  
Also, since all the neighbors of vertex~$m$ in $G^*(N_m) \bs U_A$ are in $V_1$, $i_2$ is 
not a neighbor of vertex~$m$ in $G^*(N_m) \bs U_A$, and therefore $\{m, i_2\} \not\in \tE_m$.  

Summarizing the above, we have proved that for any $i \in V_1 \cup \{m\}$ and $i_2 \in V_2$,
$\{i, i_2\} \not\in  \tE_m$.  Hence, $V_1 \cup \{m\}$ and $V_2$ are distinct components in $G^*(N_m) \bs U_A$,
so that $G^*(N_m) \bs U_A$ is disconnected.  Finally, upon noting that 
$G^*(N_{m}) \bs U_{A \cap \tX_m} = G^*(N_{m}) \bs U_A $, we see that
$A \cap \tX_{m}$ is a Type~II atom of $G^*(N_{m})$.
\endpf

%

\bigskip
When we construct $\calD_m$ by adding $\tX_m$ to $\calD_{m-1}$, 
for each Type~I atom $A$ of $G^*(N_{m-1})$, the closed curve 
$\calC_m$ is required to
\begin{list}%
{(B\arabic{cond})}{\usecounter{cond}}
\item
{\em Split} $A$ into two regions if both $A \cap \tX_m$ and $A \cap \tX_m^c$ are Type~I atoms of  $G^*(N_m)$,
so that both $A \cap \tX_m$ and $A \cap \tX_m^c$ are not suppressed in $\calD_m$;
\item
{\em Include} $A$ in $\tX_m$ if $A \cap \tX_m$ and $A \cap \tX_m^c$ are Type~I and Type~II atoms of $G^*(N_m)$,
respectively, so that $A \cap \tX_m$ is not suppressed and $A \cap \tX_m^c$ is suppressed in $\calD_m$; or
\item
{\em Exclude} $A$ from $\tX_m$ if $A \cap \tX_m$ and $A \cap \tX_m^c$ are Type~II and Type~I atoms of $G^*(N_m)$,
respectively, so that $A \cap \tX_m$ is suppressed and $A \cap \tX_m^c$ is not suppressed in $\calD_m$.
\end{list}
However, if both $A \cap \tX_m$ and $A \cap \tX_m^c$ are Type~II atoms of $G^*(N_m)$, there is
no way the closed curve $\calC_m$ can be drawn such that both of these atoms are suppressed
in $\calD_m$.  Under this situation, our recursive approach for constructing an information diagram 
for $G$ would not work.  
The following theorem (see Corollary~\r{GNm2'}) precludes this possibility.

\begin{thm}
If $A$ is a Type~I atom of $G^*(N_{m-1})$, then
\begin{list}%
{\roman{cond})}{\usecounter{cond}}
\item
if $| \, \gamma_m - U_A \, | = 0 $, then A belongs to (B3);
\item
if $| \, \gamma_m - U_A \, | = 1 $, then A belongs to (B1);
\item
if $| \, \gamma_m - U_A \, | \ge 2$, then A belongs to either (B1) or (B2).
\end{list}
\l{GNm2}
\end{thm}

\begin{cor}
If $A$ is a Type~I atom of $G^*(N_{m-1})$, 
then at least one of 
$A \cap \tX_m$ and $A \cap \tX_m^c$ is a Type~I atom of $G^*(N_m)$.
\l{GNm2'}
\end{cor}


\noindent
{\bf Proof of Theorem~\r{GNm2}} \
Let $E_m^\prime$ and $E_m^\"$ be the edge sets of $G^*(N_{m-1})$ and $G^*(N_m) \bs \{m\}$,
respectively.  
By Corollary~\r{cor3}, we have
\be
E_m^\prime = E_m^\" \cup \kappa(\gamma_m) ,
\l{wihvwe}
\ee
which implies 
\be
E_m^\prime - E_m^\" \subset \kappa(\gamma_m).
\l{2589nv}
\ee
Let $\tE_m^\prime$ and $\tE_m^\"$ be the edge sets of $G^*(N_{m-1}) \bs U_A$ and $G^*(N_m) \bs ( \{m\} \cup U_A)$,
respectively.  We see that $\tE_m^\prime$ and $\tE_m^\"$ can be obtained respectively from $E_m^\prime$ and $E_m^\"$
 by removing the edges joining at least one vertex in $U_A$ from these sets.  
Then upon removing these edges from every set in (\r{2589nv}), we obtain
\be
\tE_m^\prime - \tE_m^\" \subset \kappa ( \gamma_m - U_A ) .
\l{iavafraf}
\ee
Assume $A$ is a Type~I atom of $G^*(N_{m-1})$.  We now prove the theorem  for each of the three cases.

\noindent
i) \ $| \, \gamma_m - U_A \, | = 0$

\noindent
We first prove that  $A \cap \tX_m^c$ is a Type~I atom of $G^*(N_m)$.
Since $| \, \gamma_m - U_A \, | = 0$, i.e., $\gamma_m - U_A = \emptyset$, it follows from (\r{iavafraf})
that $\tE_m^\prime - \tE_m^\" = \emptyset$, or
$\tE_m^\prime \subset \tE_m^\"$.  In fact,
\be
\tE_m^\prime = \tE_m^\"
\l{a98phfa}
\ee
because $\tE_m^\" \subset \tE_m^\prime$ in general (cf.\ (\r{wihvwe})).
Now, if $A$ is a Type~I atom of $G^*(N_{m-1})$, then $G^*(N_{m-1}) \bs U_A$ is connected.
In view of (\r{a98phfa}), we see that $G^*(N_m) \bs ( \{m\} \cup U_A) = G^*(N_m) \bs U_{A \cap \tX_m^c}$ 
is also connected.
Therefore, $A \cap \tX_m^c$ is a Type~I atom of $G^*(N_m)$.

We now prove that $A \cap \tX_m$ is a Type~II atom of $G^*(N_m)$.
From the last paragraph, $G^*(N_m) \bs ( \{m\} \cup U_A)$ is connected.
Then all the vertices in $N_{m-1}-U_A$ are connected in $G^*(N_m) \bs ( \{m\} \cup U_A)$,
and they remain connected in $G^*(N_m) \bs U_A$ because $G^*(N_m) \bs ( \{m\} \cup U_A)$
is a subgraph of $G^*(N_m) \bs U_A$.  On the other hand, 
since $| \, \gamma_m - U_A \, | = 0$, $\{ m, i \}$ is not an edge in $G^*(N_m)$ for all $i \in N_{m-1} - U_A$,
and so in in $G^*(N_m)$,
vertex~$m$ is not connected with any vertex in $N_{m-1}-U_A$.  Since $G^*(N_m) \bs U_A$ is a subgraph of
$G^*(N_m)$, we see that in $G^*(N_m) \bs U_A$,
vertex~$m$ is not connected with any vertex in $N_{m-1}-U_A$.
Thus $G^*(N_m) \bs U_A$ is disconnected.
Hence, we conclude that $G^*(N_m) \bs U_{A  \cap \tX_m} = G^*(N_m) \bs U_A$ is disconnected, i.e.
$A \cap \tX_m$ is a Type~II atom of $G^*(N_m)$.

%
%

\noindent
ii) \ $| \, \gamma_m - U_A \, | = 1$

\noindent
To prove that $A \cap \tX_m^c$ is a Type~I atom of $G^*(N_m)$, we only have to observe that 
in (\r{iavafraf}), $\kappa ( \gamma_m - U_A ) = \emptyset$ when $| \, \gamma_m - U_A \, | = 1$.
Then (\r{a98phfa}) holds and we can apply the same argument as in case~i).

We now prove that $A \cap \tX_m$ is a Type~I atom of $G^*(N_m)$.
For the economy of presentation, we will give a proof that also covers case~iii), i.e., $| \, \gamma_m - U_A \, | \ge 2$.

It follows from (\r{iavafraf}) that for $i, j \in N_{m-1} - U_A$, if $\{i,j\} \in \tE_m^\prime$ and $\{i,j\} \not\in \tE_m^\"$, then 
$\{i, j\} \in \kappa ( \gamma_m - U_A )$, i.e., $i, j \in \gamma_m - U_A$.  Equivalently,
if $i \not\in \gamma_m - U_A$ or $j \not\in \gamma_m - U_A$, then $\{i,j\} \not\in \tE_m^\prime$
or $\{i,j\} \in \tE_M^\"$. In other words, if $i$ and $j$ are not both in $\gamma_m - U_A$
and $\{i,j\} \in \tE_m^\prime$, then $\{i,j\} \in \tE_m^\"$.  

Since $A$ is a Type~I atom of $G^*(N_{m-1})$, $G^*(N_{m-1}) \bs U_A$ is connected.
Then for any distinct $\alpha, \beta \in N_{m-1} - U_A$,
there exists a path between vertices $\alpha$ and $\beta$ in $G^*(N_{m-1}) \bs U_A$.

If such a path does not contain an edge with both endpoints in $\gamma_m - U_A$,
from the discussion in the second last paragraph, we see that all the edges on 
the path are in 
$G^*(N_m) \bs ( \{m\} \cup U_A)$ and hence are also in 
$G^*(N_m) \bs U_A$ because $G^*(N_m) \bs ( \{m\} \cup U_A)$ is a subgraph of 
$G^*(N_m) \bs U_A$.  In other words, 
$\alpha$ and $\beta$ are connected in $G^*(N_m) \bs U_{A}$.

If the path contains any edge with both endpoints in $\gamma_m - U_A$, we can construct 
another path between $\alpha$ and $\beta$ in $G^*(N_m) \bs U_{A}$ by replacing every such 
an edge $\{u,v\}$ by edges $\{u,m\}$ and $\{m,v\}$ which are both in $G^*(N_m) \bs U_A$.  Then we see that 
$\alpha$ and $\beta$ are connected in $G^*(N_m) \bs U_{A}$.

Finally, since $\gamma_m \not\subset U_A$, there exists a vertex $u \in \gamma_m - U_A$.
Then for any $\alpha \in N_{m-1} - U_A$, we can readily see that vertices $\alpha$ and $m$ are connected
in $G^*(N_m) \bs U_{A}$ because $\alpha$ is connected with $u$ and $u$ is connected with $m$.
Summarizing the above, for any distinct $\alpha, \beta \in N_m - U_A$, $\alpha$ and $\beta$ are connected
in $G^*(N_m) \bs U_{A}$.  This implies that $G^*(N_m) \bs U_{A} = G^*(N_m) \bs U_{A \cap \tX_m}$
is connected, i.e., $A \cap \tX_m$ is a Type~I atom of $G^*(N_m)$.

\noindent
ii) \ $| \, \gamma_m - U_A \, | \ge 2$

\noindent
To prove that $A$ belongs to either (B1) or (B2), we only need to prove that
$A \cap \tX_m$ is a Type~I atom of $G^*(N_m)$. This has already been proved in case~ii).
\endpf

\bigskip
Now in light of Theorem~\r{GNm} and Corollary~\r{GNm2'}, we can construct $\calD_m$ from $\calD_{m-1}$ 
according to the prescriptions in (B1)-(B3) for each Type~I atom of  $G^*(N_{m-1})$.  Theorem~\r{GNm2} helps 
simplify the checking of which of (B1) to (B3) these atoms belong to, as we now explain.  The three conditions for $A$ in Theorem~\r{GNm2} are
\begin{list}%
{\roman{cond})}{\usecounter{cond}}
\item
$| \, \gamma_m - U_A \, | = 0 $: This means that $A \not\subset \tX_i$ for any $i \in \gamma_m$, or equivalently, $A \not\subset \tX_{\gamma_m}$.  In this case, the closed curve $\calC_m$ excludes the atom~$A$.
\item
$| \, \gamma_m - U_A \, | = 1 $: This means that $A \subset \tX_i$ for only one $i \in \gamma_m$.
In this case, the closed curve $\calC_m$ splits the atom~$A$.
\item
$| \, \gamma_m - U_A \, | \ge 2 $: This means that $A \subset \tX_i$ for more than one $i \in \gamma_m$.
If $A$ belongs to (B1), the closed curve $\calC_m$ splits the atom~$A$, otherwise ($A$ belongs to
(B2)) it includes the atom $A$.
\end{list}
Note that if $A$ satisfies condition~iii), we still need to check whether it belongs to (B1) or (B2),
which is unavoidable.  
This is explained by the two examples in Figs.~\r{fig12} and~\r{fig13}.
In each of these examples, we have $n = 3$ and $\gamma_3 = \{1,2\}$.  Consider the Type~I atom $A = 12$ of $G^*(N_2)$.
Then $U_A = \emptyset$ and $| \, \gamma_3 - U_A \, | = 2$, i.e., $A$~satisfies condition~iii).

\begin{eg}
We apply our method to construct an information diagram for the Markov chain 
$X_1 \ra X_2 \ra \cdots \ra X_n$.  From Corollary~\r{q5;once}, we see that for $m = n, n-1, \cdots, 1$,
$G^*(N_m)$ is the path $P_n$ defined in Section~\r{sec5}. 

The information diagram $\calD_1$, which consists of only the set variable $\tX_1$, is completely trivial.
For $2 \le m \le n$, when we construct $\calD_m$ from $\calD_{m-1}$, since $\gamma_m = \{m-1\}$
(cf.\ (\r{gammam})) is a singleton, every Type~I atom $A \subset \tX_{m-1}$ in $\calD_{m-1}$
satisfies the condition that $A \subset \tX_i$ for only one $i \in \gamma_m$.
Then based on the discussion in the foregoing, 
the closed curve $\calC_m$ excludes every Type~I atom in $\calD_{m-1}$
that is not contained in
$\tX_{m-1}$ and splits every Type~I atom in $\calD_{m-1}$ that is contained in $\tX_{m-1}$.
This way, we can obtain the information diagram in Fig.~\r{figMC}.
\l{eg16}
\end{eg}

Our method is robust in the sense that it works for any labeling of the vertices.
However, it may be more convenient to construct the information diagram for one labeling than another.  
This is illustrated in the following two examples.

\begin{eg}
Consider the graph $G$ in Fig.~\r{figstar}.  Upon noting that $\gamma_3$ and $\gamma_4$ are singletons
(both equal to $\{2\}$),
by using the technique in Example~\r{eg16},
we can readily construct the information diagram in Fig.~\r{diagstar1}. 
\l{eg17}
\end{eg}

\begin{eg}
Consider the graph $G$ in Fig.~\r{figeg11}, which is the same as the one in Fig.~\r{figstar} except that 
the vertices are labelled differently.  Here, $G^*(N_3)$ is the complete graph $K_3$, and so 
there is no Type~II atom in $\calD_3$.  Now $\gamma_4 = \{ 1, 2, 3 \}$.  We have to consider
all the atoms that are subsets of $\tX_{\gamma_4} = \tX_1 \cup \tX_2 \cup \tX_3$, 
namely all the atoms in $\calD_3$ because 
$N_3 - \gamma_4 = \emptyset$.
For the atoms
$1\bar{2}\bar{3}$, $\bar{1}2\bar{3}$, and $\bar{1}\bar{2}3$ in $\calD_3$, since each
of them is a subset of $\tX_i$ for only one $i \in \gamma_4$, the closed curve 
$\calC_4$ splits each of these atoms.  For an atom $A$ equal to $\bar{1}23$, $1\bar{2}3$, or $12\bar{3}$, 
we can check that $A \cap \tX_4 \in \TI(G)$ and $A \cap \tX_4^c \in \TII(G)$.  Therefore, $A$ 
belongs to (B2).  For the atom $123$, we can check that $1234 \in \TI(G)$ and $123\bar{4}  \in \TII(G)$, 
so this atom also belongs to (B2).  As such, the closed curve $\calC_4$ includes all these 
four atoms, and we can construct the information diagram in Fig.~\r{figeg18}.  It can readily to checked
that this information diagram is equivalent to the one in Fig.~\r{diagstar1} constructed in Example~\r{eg17}.
\l{eg18}
\end{eg}

\begin{eg}
Figs.~\r{figeg19} and \r{figeg19b} show a graph $G$ and the corresponding information diagram, respectively.
In light of Corollary~\r{q5;once}, we see that $G^*(N_4)$ is the path $P_4$ defined 
in Section~\r{sec5}. Then we can use the information diagram for $X_1 \ra X_2 \ra X_3 \ra X_4$ 
as $\calD_4$ and add to it the closed curves $\calC_5$ and $\calC_6$ using the 
technique in Example~\r{eg16} to obtain the information diagram in Fig.~\r{figeg19b}.
\l{eg19}
\end{eg}



\section{Conclusion}
\l{sec7}
The theory of $I$-Measure proves to be a very useful tool for characterizing
full conditional independence structures and MRFs \cite{YeungLY02},
because with the $I$-Measure, the fundamental set-theoretic structure of the problem is revealed.
In this paper, we apply this tool to obtain three main results related to MRFs.

For an MRF represented by an undirected graph, a subfield is a subset of the random variables forming the MRF .
We have determined the smallest undirected graph 
that can always represent a subfield as an MRF.  
This is our first main result.
As an application of this result, we have obtained a necessary and sufficient condition for a
subfield of a Markov tree to be also a Markov tree.

A Markov chain can be regarded a special case of an MRF.  It was previously known that 
the $I$-Measure of a Markov chain is always nonnegative \cite{KawabataY92}.  Here, we have proved
that the $I$-Measure is nonnegative
for every MRF represented by a given undirected graph 
if and only if the graph is a forest of paths, i.e., the Markov random 
field is a collection of independent Markov chains.  This means that
Markov chains are essentially the only MRFs such that the $I$-Measure is always nonnegative.
This is our second main result.  In the course of proving this result, we have obtained
some interesting properties of the $I$-Measure pertaining to an MRF.

Our third main result is a nontrivial application of our first main result.
In \cite{KawabataY92}, a construction of an information diagram for a Markov chain
was presented.  By applying our first main result, 
we have developed
a recursive approach for constructing information diagrams for MRFs. 
Such diagrams not only reveal the special structure of the $I$-Measure for an MRF, 
but they also are very useful for identifying information identities and 
inequalities pertaining to an MRF.

The work in our paper is based upon the the view that an MRF is a collection of 
full conditional mutual independencies.  As such, some of our results can potentially
be generalized for Markov structures beyond MRFs.



\appendix

\section{Proof of Proposition~\r{V"}}
\l{**}
Let 
\ba
G^*(V') &=& (V', E'),  \\
G^*(V^\") &=& (V^\", E^\") ,
\ea
and 
\[
(G^*(V'))^*(V^\") = (V^\", \tE^\").
\]
Consider any $\{u,v\} \in E^\"$.  By the definition of $G^*(V^\")$, there exists a path between $u$ and $v$ in $G$ on which all the intermediate vertices 
are in 
$V-V^\"$. Now on this path, let $w_1, w_2, \cdots, w_k$ be the vertices in $V' - V^\"$ in the direction from $u$ to $v$.
Then the vertices between $u$ and $w_1$, the vertices between $w_i$ and $w_{i+1}$ for $1 \le i \le k-1$, and 
the vertices between $w_k$ and $v$ are all in $(V-V^\") - (V'-V^\") = V-V'$, because $V^\" \subset V' \subset V$.  By the definition of $G^*(V')$, the edges 
$\{u, w_1\}$, $\{w_i, w_{i+1}\}$ for $1 \le i \le k-1$,
and $\{w_k, v\}$ are in $E'$.  In other words, there is a path between $u$ and $v$ in $G^*(V')$
on which all the intermediate vertices are in $V' - V^\"$.
Then by the definition of $(G^*(V'))^*(V^\")$, $\{u,v\} \in \tE^\"$. 
This proves that $E^\" \subset \tE^\"$.

On the other hand, consider any $\{u,v\} \in \tE^\"$.  By the definition of $(G^*(V'))^*(V^\")$,
there exists a path between $u$ and $v$ in $G^*(V')$ on which all the intermediate vertices are in $V' - V^\"$.  
Let the edges on this path be $\{u, w_1\}$, $\{w_1, w_2\}$, $\cdots$, $\{w_{k-1}, w_k\}$,
and $\{w_k, v\}$, where $w_i, 1 \le i \le k$ are in $V' - V^\"$.  Since $\{u, w_1\} \in E'$, by the definition of $G^*(V')$,
there exists a path between $u$ and $w_1$ on which all the intermediate nodes are in $V-V'$.
Similarly, there exists a path between 
$w_i$ and $w_{i+1}$ for $1 \le i \le k$, and a path between $w_k$ and $v$,
on which all the intermediate nodes are in $V-V'$.  Thus there exists
a path between $u$ and $v$ in $G$ on which all the intermediate nodes are in 
$(V-V') \cup (V'-V^\") = V-V^\"$, because $V^\" \subset V' \subset V$.  
Then by the definition of $G^*(V^\")$, $\{u,v\} \in E^\"$.
This proves that $\tE^\" \subset E^\"$.

Hence, we conclude that $\tE^\" = E^\"$, i.e., $(G^*(V'))^*(V^\") = G^*(V^\")$.
The proposition is proved.

\section{Verification of $\mu^*$ in the Proof of Theorem~\r{thm10}}
\l{App_A}
In this appendix, we verify that 
$\mu^*$ as prescribed by (\r{98hn;oat}) satisfies (\r{q89uy5}).
First, for $i \in V$, consider
\ba
\mu^*(\tX_i) & = & \mu^* \left( \bigcup_{S \subset V - \{i\}} \left( \tX_i \cap \tX_S - \tX_{V-S-\{i\}} \right) \right) \\
& = & \sum_{S \subset V}  \mu^* \left( \tX_i \cap \tX_S - \tX_{V-S-\{i\}} \right) .
\ea
From (\r{98hn;oat}), we see that $\mu^*(\cdot)$ in the above summation vanishes if $|S| \le n-3$, and so
\begin{eqnarray}
\mu^*(\tX_i) & = & \sum_{j \ne i} \mu^* 
\left( \bigcap_{k \ne j} \tX_k - \tX_j \right) + \mu^* \left( \tX_0 \cap 
\tX_1 \cap \cdots \cap \tX_{n-1} \right) \nn \\
& = & (n-1) \cdot 1- (n-2) \nn \\
& = & 1.
\l{98yggs}
\end{eqnarray}
This verifies (\r{q89uy5}) for the case $|B|=1$.
Next, for $0 \le i < j \le n-1$, consider 
\begin{eqnarray}
\mu^* ( \tX_i \cap \tX_j ) 
& = & \sum_{S \subset V - \{i,j\}} \mu^* \left( \tX_i \cap \tX_j \cap \left( \bigcap_{k \in S} \tX_k \right)
- \left( \bigcup_{l \in V - \{i,j\} - S} \tX_l \right) \right) \nn \\
& = & \sum_{S \subset V - \{i,j\} \atop |S| \ge n-3 } \mu^* \left( \tX_i \cap \tX_j \cap \left( \bigcap_{k \in S} \tX_k \right)
- \left( \bigcup_{l \in V - \{i,j\} - S} \tX_l \right) \right) \nn \\
& = & \sum_{m \ne i,j} \mu^* \left( \bigcap_{r \in V - \{m\}} \tX_r - \tX_m \right) + \mu^* \left( \bigcap_{s=1}^n
\tX_s \right) \nn \\
& = & (n-2) \cdot 1 - (n-2) \nn \\
& = & 0.
\l{mnvfmn}
\end{eqnarray}
It follows from (\r{98yggs}) and (\r{mnvfmn}) that
\begin{eqnarray}
\mu^* ( \tX_i \cup \tX_j )
& = & \mu^* (\tX_i ) + \mu^* (\tX_j ) - \mu^* ( \tX_i \cap \tX_j ) \\
& = & 1 + 1 - 0 \\
& = & 2.
\l{anq4u}
\end{eqnarray}
This verifies (\r{q89uy5}) for the case $|B|=2$.  Now for $1 \le i < j < k \le n$, consider 
\ba
\lefteqn{ \mu^* ( \tX_k - ( \tX_i \cup \tX_j )) } \\
& = & \sum_{S \subset V - \{i,j,k\}} \mu^* \left( \tX_k \cap \left( \bigcap_{l \in S} \tX_l \right) -
\left( \tX_i \cup \tX_j \cup \bigcup_{m \in V - \{i,j,k\} -S} \tX_m \right) \right) .
\ea
In the above, since $| k \cup S | \le n-2$, we see from (\r{98hn;oat}) that every term in the above summation
vanishes, and so
\be
\mu^* ( \tX_k - ( \tX_i \cup \tX_j )) = 0.
\l{fjvar5}
\ee
Finally, consider $B \subset V$ such that $3 \le |B| \le n$ and let $i, j$ be two arbitrary elements of $B$.
Then in light of (\r{anq4u}) and (\r{fjvar5}), we have
\ba
\mu^*(\tX_B) 
& = & \mu^* ( \tX_i \cup \tX_j ) + \mu^* \left( \tX_{B-\{i,j\}} - ( \tX_i \cup \tX_j ) \right) \\
& \le & \mu^* ( \tX_i \cup \tX_j ) + \sum_{k \in B-\{i,j\}} \mu^* ( \tX_k - ( \tX_i \cup \tX_j ) ) \\
& = & 2 + 0 \\
& = & 2,
\ea
where the inequality above is justified by the union bound because $\mu^*$ is nonnegative on all the atoms in
$\tX_{B-\{i,j\}} - ( \tX_i \cup \tX_j )$ (cf.\ (\r{98hn;oat})). On the other hand, we have
\ba
\mu^*(\tX_B) 
& = & \mu^* ( \tX_i \cup \tX_j ) + \mu^* \left( \tX_{B-\{i,j\}} - ( \tX_i \cup \tX_j ) \right) \\
& \ge & \mu^* ( \tX_i \cup \tX_j ) \\
& = & 2,
\ea
again because $\mu^*$ is nonnegative on all the atoms in
$\tX_{B-\{i,j\}} - ( \tX_i \cup \tX_j )$.
Therefore, $\mu^*(\tX_B)  = 2$,
verifying (\r{q89uy5}) for the case $3 \le |B| \le n$.

\section{Proof of Lemmas~\r{vanetl} and \r{kffgjlj}}
\l{App_B}

In this appendix, we prove Lemmas~\r{vanetl} and \r{kffgjlj}
via the following lemma.

\begin{lemma}
Let $G = (V,E)$ be a connected undirected graph and $B = \{ v \in V : s(\{v\}) = 1 \}$.  Then
\begin{list}%
{\alph{cond})}{\usecounter{cond}}
\item
for any $k \in V-B$, we have $B \cap J_i \ne \emptyset$ for all $1 \le i \le s_k$, where $J_1, J_2, \cdots, J_{s_k}$ 
  ($s_k \ge 2$) are the components of $G \bs \{k \}$;
\item
$s(R) > 1$ for all nonempty subset $R$ of $V - B$.
\end{list}
\l{auinfv}
\end{lemma}

\pf
We assume that $B \ne V$, since otherwise $V- B = \emptyset$ and the lemma has no assertion.

We first prove a).  
Let $k \in V- B$, and by the definition of $B$, we have $s_k \ge 2$.
Consider any spanning tree $T$ of $G$.  Note that $T$ must contain 
at least one edge connecting $k$ and each $J_i$ ($1 \le i \le s_k$) because $G$ is connected.
For any fixed $i$, consider such an edge and call it $e$. 
Upon removing $e$, $T$ is disconnected with one component being a subtree
containing $k$ and the other component being a subtree not containing $k$. 
For the latter subtree, all the vertices are in $J_i$, otherwise there exists an edge connecting $J_i$ 
and $J_{i'}$ ($i' \ne i$), which is a contradiction 
because $J_1, J_2, \cdots, J_{s_k}$ are the components of $G \bs \{k \}$.  Then this subtree must have at least one
leaf in $J_i$, say $l$.  Note that $\{ l \}$ is not a cutset in $T$ and hence not a cutset in 
$G$.  Therefore $B \cap J_i \ne \emptyset$, proving a).

We now prove b).  Consider any nonempty subset $R$ of $V - B$ and fix $k \in R$.
Since $R \subset V - B$, we have $R \cap B = \emptyset$.
By a), $B \cap J_i \ne \emptyset$ for all $1 \le i \le s_k$, with $s_k \ge 2$.
Then the vertices in $B$ are not all connected in $G \bs \{ k\}$, and hence
not all connected in $G \bs R$ because $k \in R$ and $R \cap B = \emptyset$.  
This implies that $G \bs R$ is not connected, or $s(R) > 1$.  The lemma is proved.
\endpf

\bigskip
Lemmas~\r{vanetl} and \r{kffgjlj} can now be obtained as follows.
In Theorem~\r{kfklaelr}, $U_A$ is a Type~I atom, and so $G \bs U_A$ is connected.
The same holds for Lemmas~\r{vanetl} and \r{kffgjlj}.
Lemma~\r{vanetl} is trivial for $|U_A| = n-2$.  For $|U_A| < n-2$, it can be obtained by  
applying Part a) of Lemma~\r{auinfv} to $G \bs U_A$. 
Finally, by applying Part b) of Lemma~\r{auinfv} to $G \bs U_A$ with $R = W - S$, we obtain Lemma~\r{kffgjlj}.

\bigskip
\bigskip
\noindent
{\Large \bf Acknowledgment} 

\medskip
\noindent
The authors would like to thank Prof.\ Leizheng Cai, and Prof.\ Franti\v{s}ek Mat\'{u}\v{s}
for the useful discussions.

\newpage

\begin{figure}[h]
\vspace{2in}
\centering
\includegraphics[height=3in]{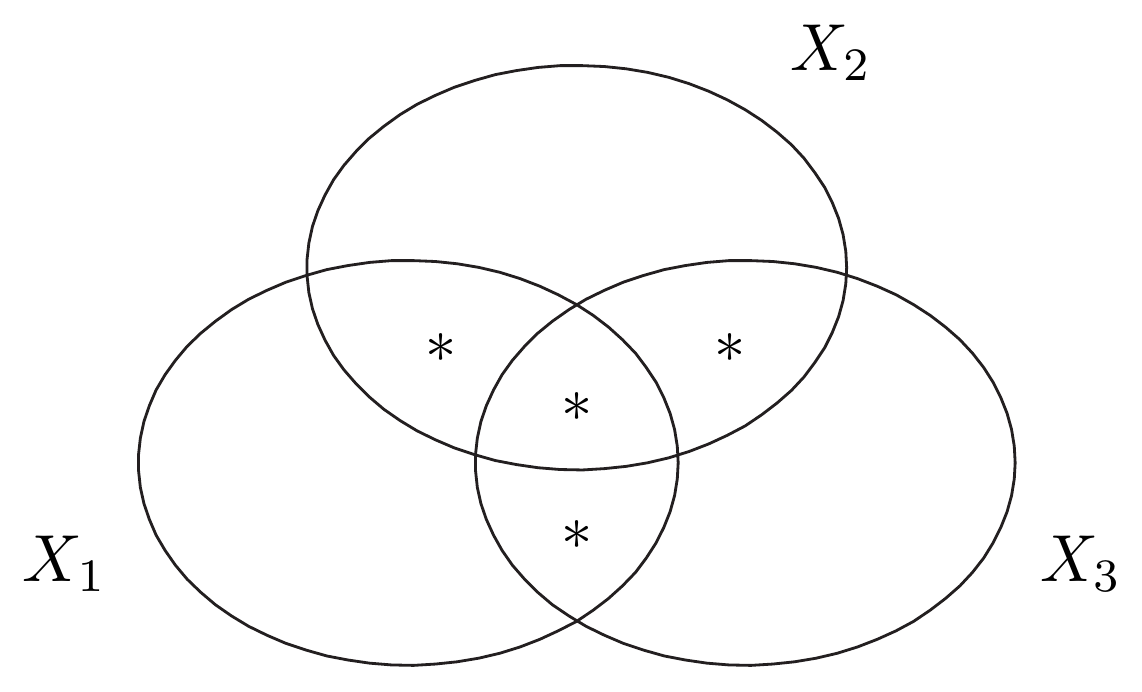}
\caption{The information diagram for Example~\r{eg2}.}
\l{figeg2}
\end{figure}

\newpage

\begin{figure}[h]
\vspace{2in}
\centering
\includegraphics[height=3in]{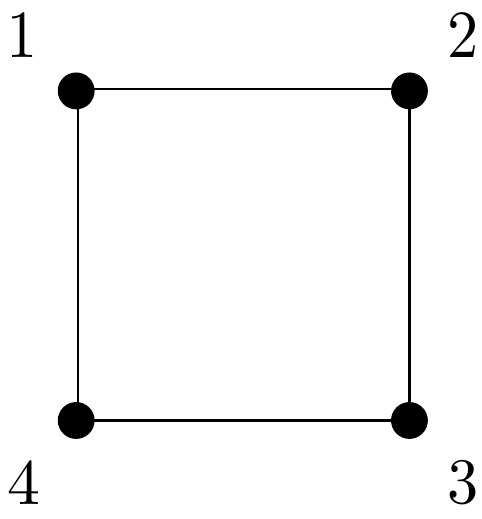}
\caption{The graph $G$ in Example~\r{eg5}.}
\l{figeg5}
\end{figure}

\newpage

\begin{figure}[h]
\vspace{2in}
\centering
\includegraphics[height=2.5in]{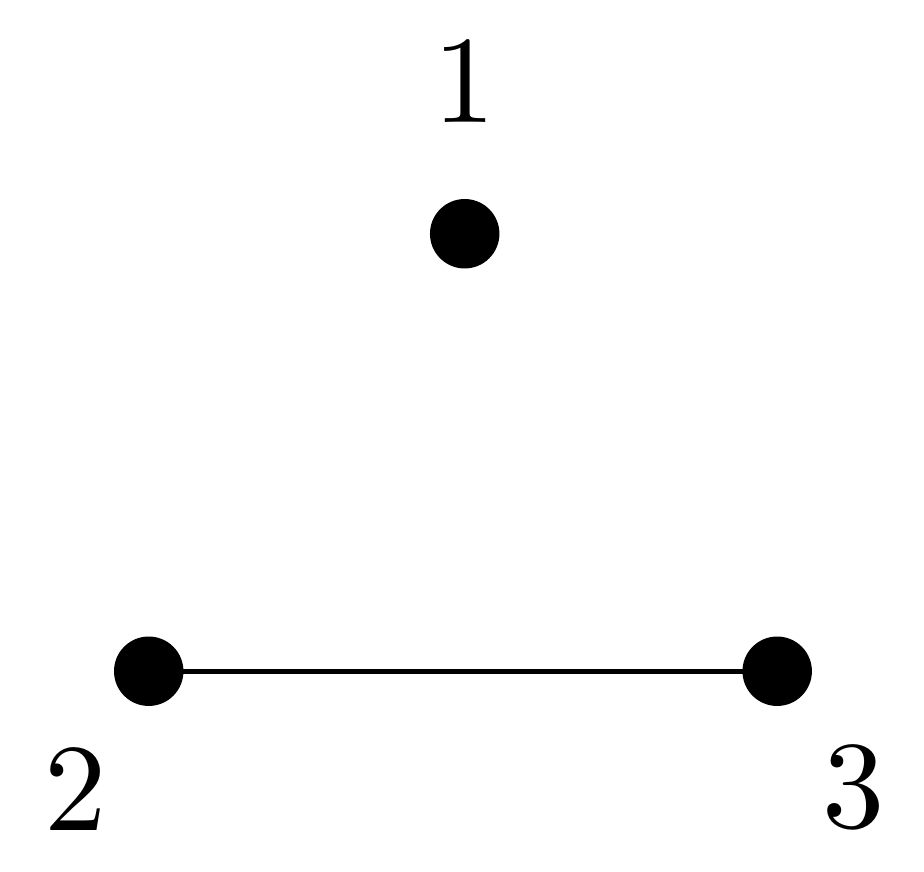}
\caption{The graph $\hat{G}$ corresponding to the set ${\cal A}_{{\rm II}}$ in Example~\r{eg6}.}
\l{figeg6}
\end{figure}

\newpage

\begin{figure}[h]
\vspace{3in}
\centering
\includegraphics[width=5.5in]{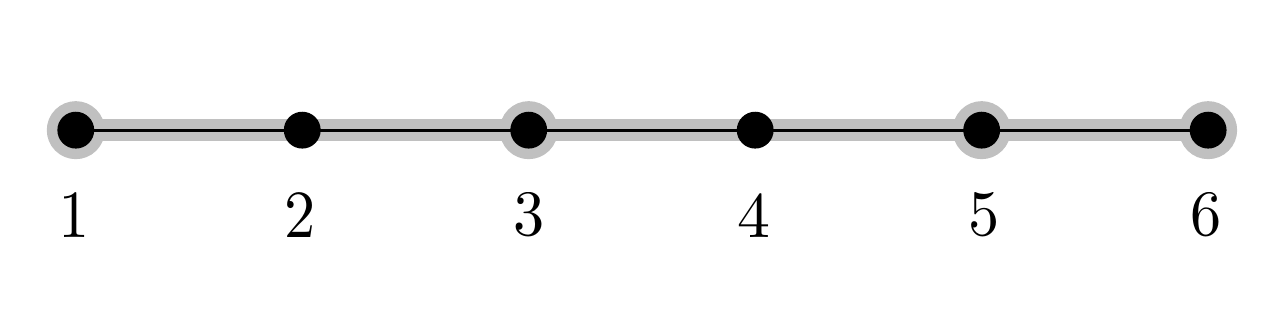}
\caption{The graphs $G$ (black) and $G^*(V')$ (grey) in Example~\r{eg7},
with $V' \{ 1, 3, 5, 6 \}$.  }
\l{figeg7}
\end{figure}

\newpage

\begin{figure}[h]
\vspace{1.5in}
\centering
\includegraphics[height=4in]{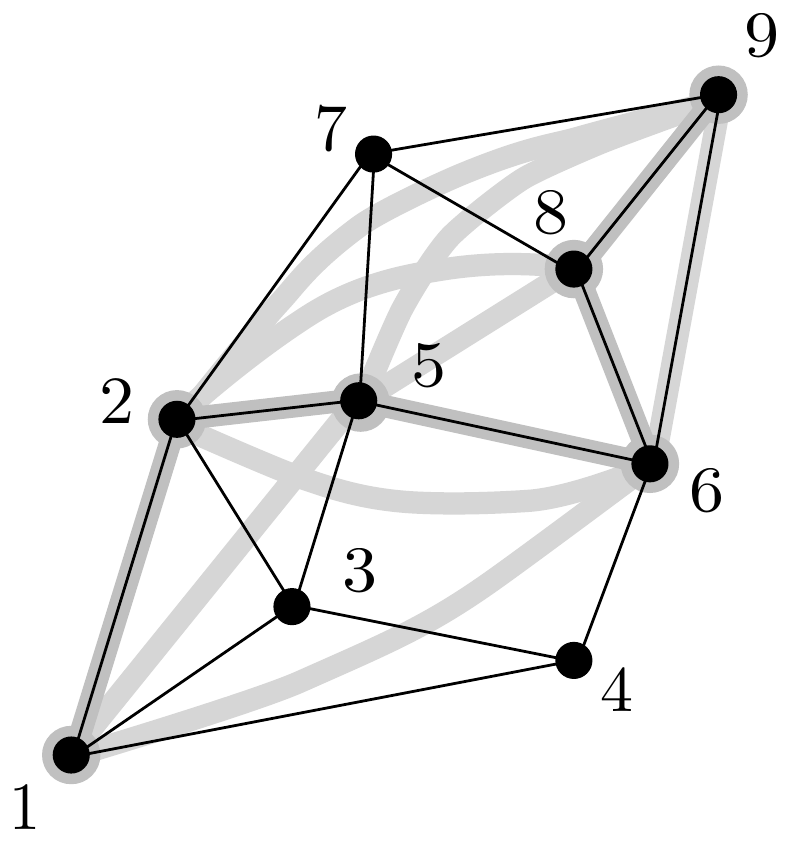}
\caption{The graphs $G$ (black) and $G^*(V')$ (grey) in Example~\r{eg8}, with $V' = \{ 1, 2, 5, 6, 8,  9 \}$.}
\l{figeg8}
\end{figure}

\newpage

\begin{figure}[h]
\vspace{2in}
\centering
\includegraphics[height=3in]{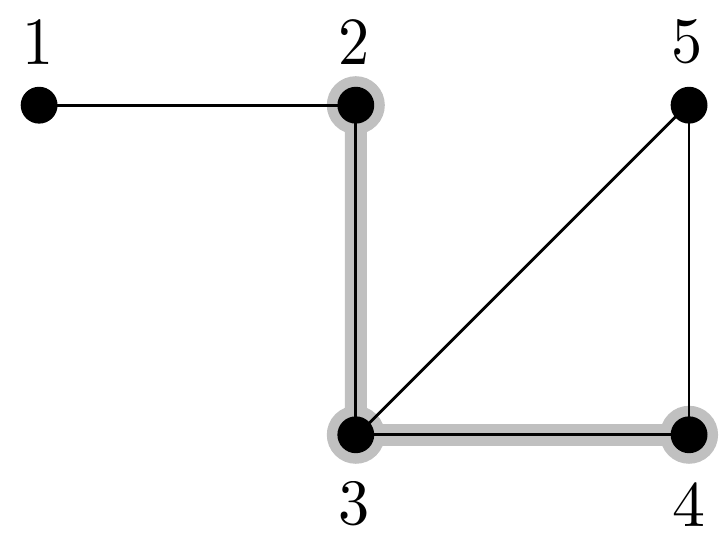}
\caption{The graphs $G$ (black) and $G^*(V')$ (grey) in  Example~\r{eg9}, with $V' = \{ 2, 3, 4 \}$.}
\l{figeg9}
\end{figure}

\newpage

\begin{figure}[h]
\vspace{2in}
\centering
\includegraphics[height=3.5in]{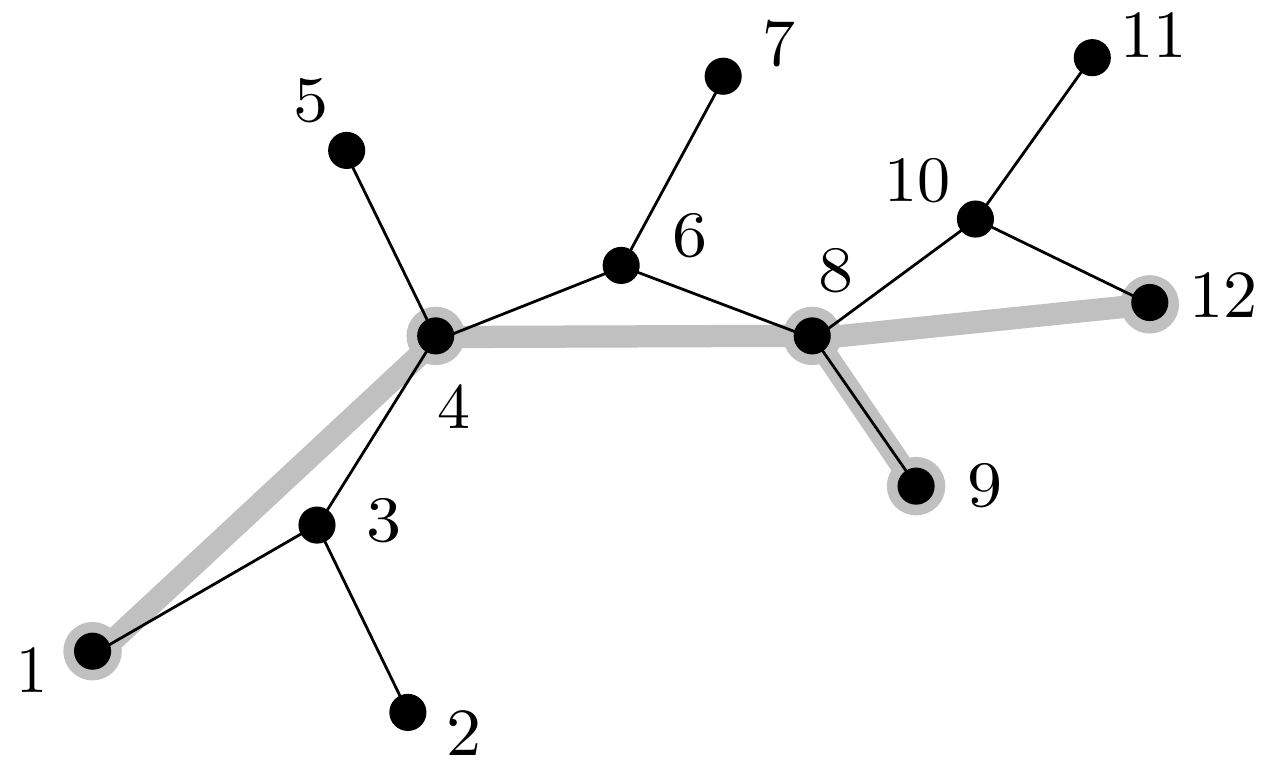}
\caption{The trees $G$ (black) and $G^*(V')$ (grey) in Example~\r{eg10}, with
$V' = \{1, 4, 8, 9, 12 \}$.}
\l{figeg10}
\end{figure}

\newpage

\begin{figure}[h]
\vspace{2in}
\centering
\includegraphics[height=3.5in]{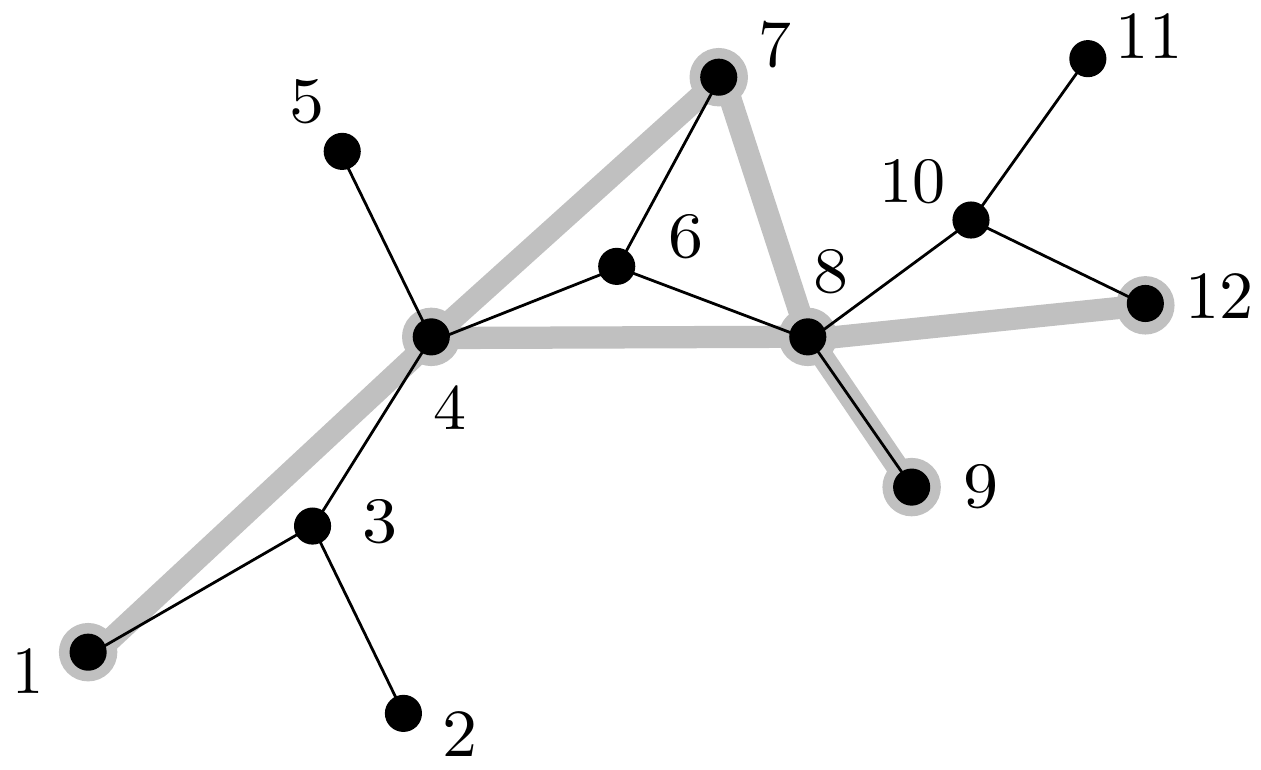}
\caption{The tree $G$ (black) and the graph $G^*(V')$ (grey) in Example~\r{eg10}, with
$V' = \{1, 4, 7, 8, 9, 12 \}$.}
\l{figeg10b}
\end{figure}

\newpage

\begin{figure}[h]
\vspace{2in}
\centering
\includegraphics[height=2.25in]{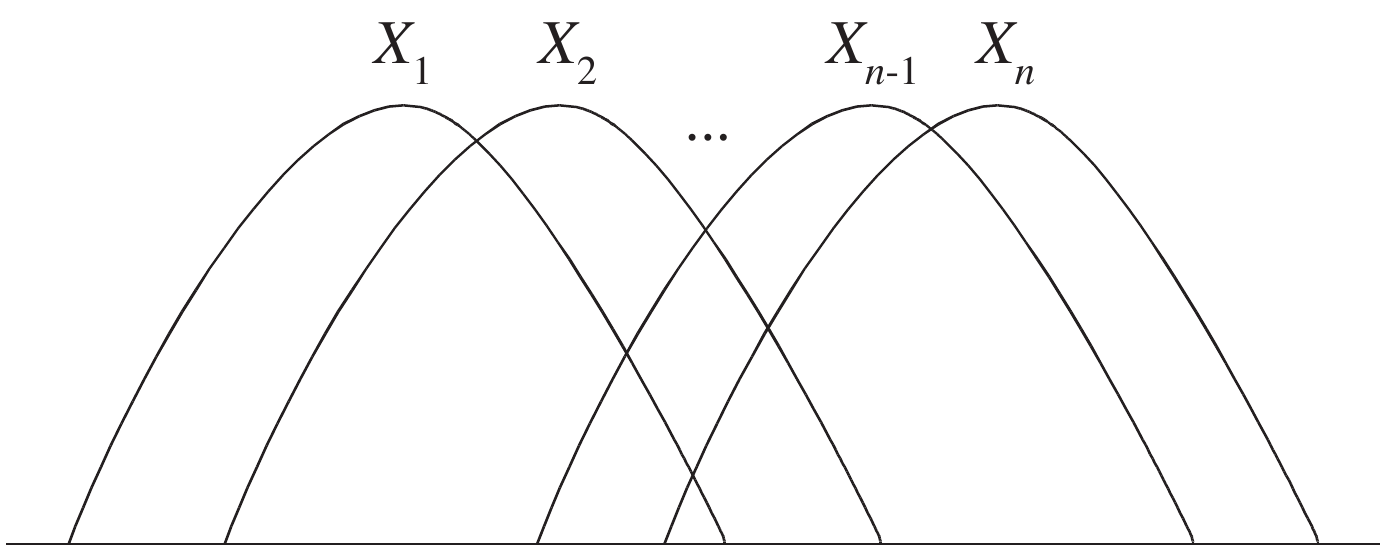}
\caption{The information diagram for the Markov chain $X_1 \ra X_2 \ra \cdots \ra X_n$.}
\l{figMC}
\end{figure}

\newpage

\begin{figure}[h]
\vspace{1in}
\centering
\includegraphics[height=3in]{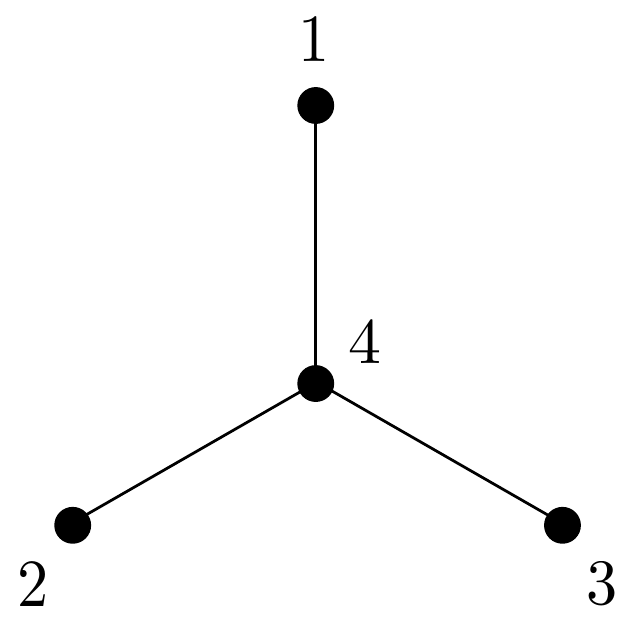}
\caption{The ``star" representing the Markov tree in Example~\r{eg11}.}
\l{figeg11}
\end{figure}

\newpage

\begin{figure}[h]
\vspace{2.5in}
\centering
\includegraphics[height=2.5in]{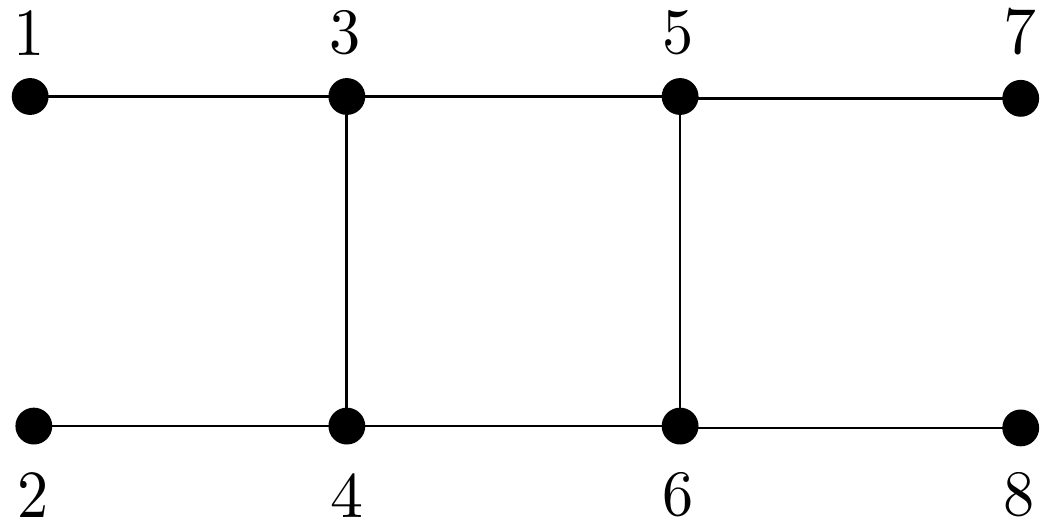}
\caption{The graph in Example~\r{eg12}.}
\l{figeg12}
\end{figure}

\newpage

\begin{figure}[h]
\vspace{2.5in}
\centering
\includegraphics[height=2.75in]{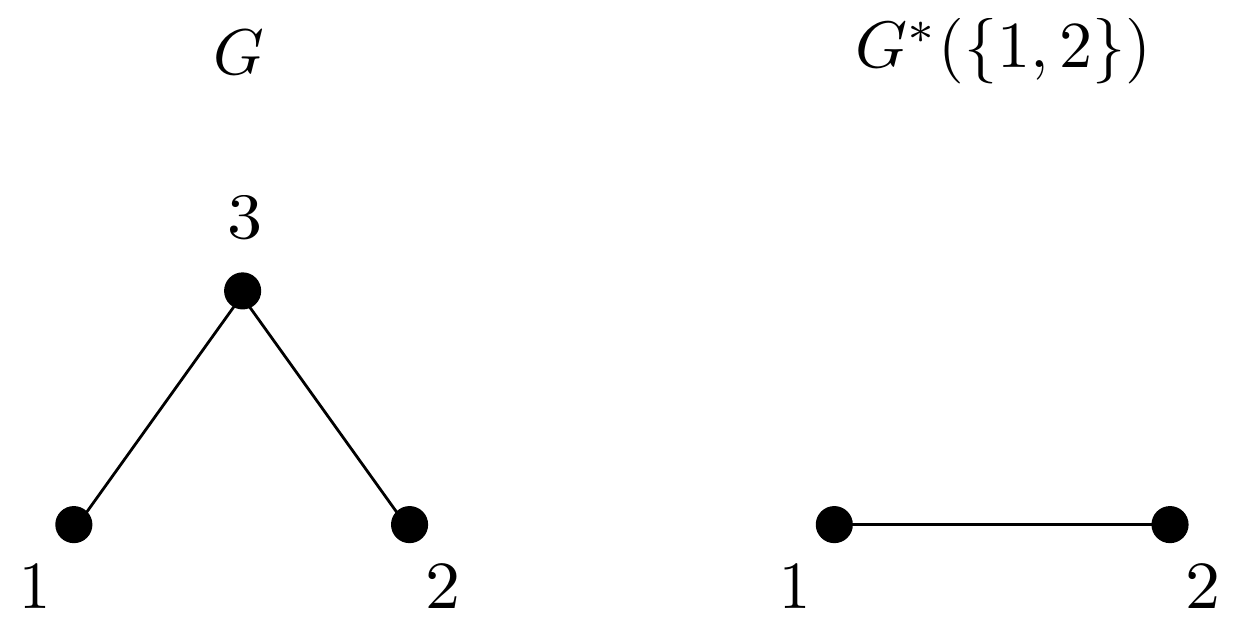}
\caption{In this example, $A = 12 \in \TI(G^*(\{1,2\}))$, $A \cap \tX_3 = 123 \in \TI(G)$, and $A \cap \tX_3^c = 12\bar{3} \in \TII(G)$. Therefore, $A$ belongs to (B2).}
\l{fig12}
\end{figure}

\newpage

\begin{figure}[h]
\vspace{2.5in}
\centering
\includegraphics[height=2.75in]{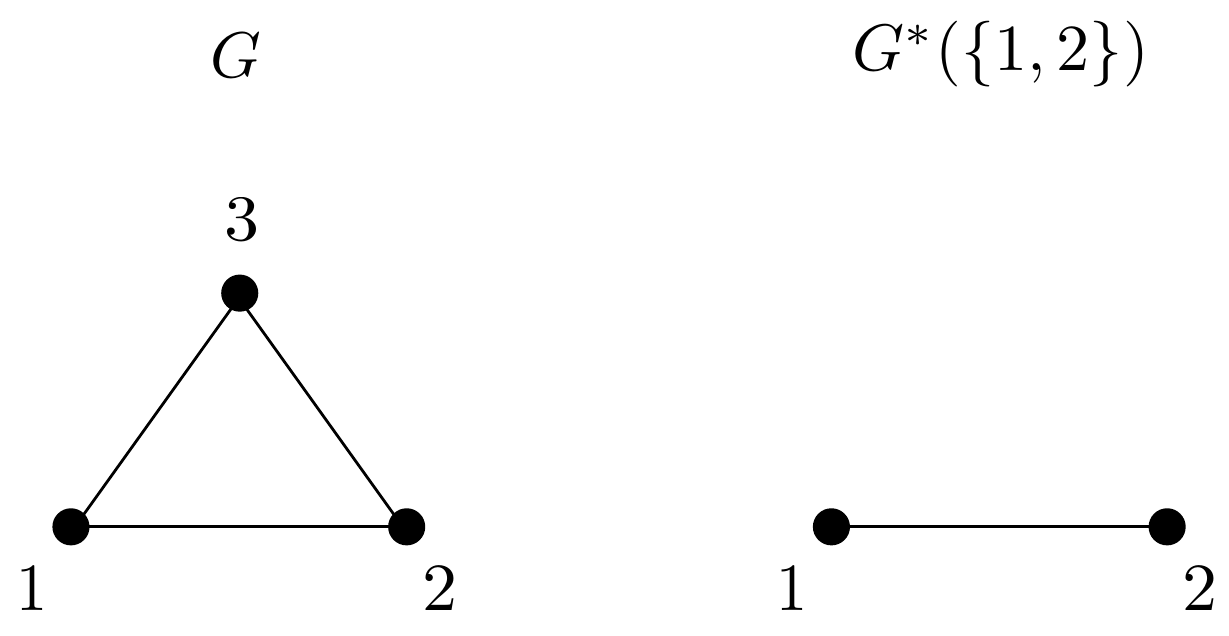}
\caption{In this example, $A = 12 \in \TI(G^*(\{1,2\}))$, $A \cap \tX_3 = 123 \in \TI(G)$, and $A \cap \tX_3^c = 12\bar{3} \in \TI(G)$. Therefore, $A$ belongs to (B1).}
\l{fig13}
\end{figure}

\newpage

\begin{figure}[h]
\vspace{2.5in}
\centering
\includegraphics[height=3in]{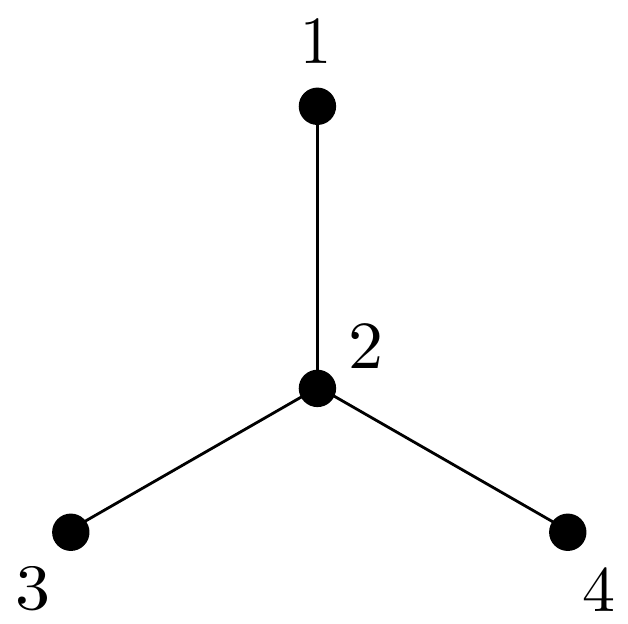}
\caption{The graph $G$ in Example~\r{eg17}.}
\l{figstar}
\end{figure}

\newpage

\begin{figure}[h]
\vspace{2.5in}
\centering
\includegraphics[height=3.25in]{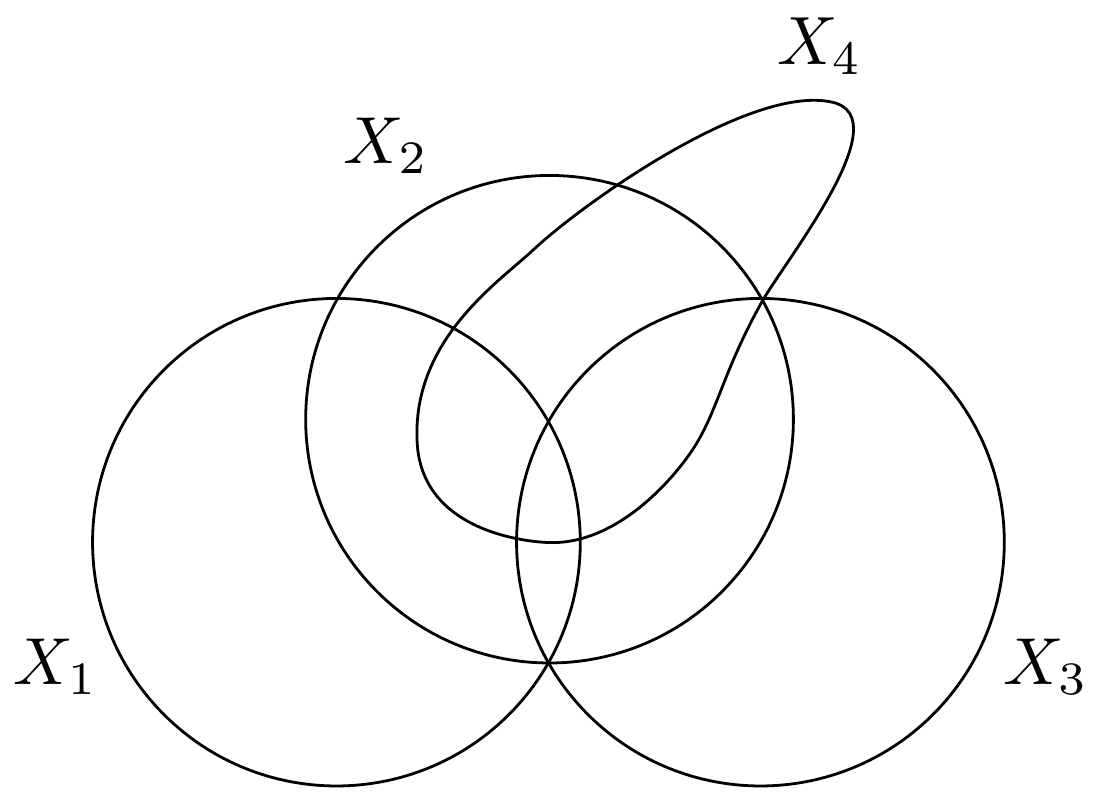}
\caption{The information diagram in Example~\r{eg17}.}
\l{diagstar1}
\end{figure}

\newpage

\begin{figure}[h]
\vspace{2.5in}
\centering
\includegraphics[height=2.75in]{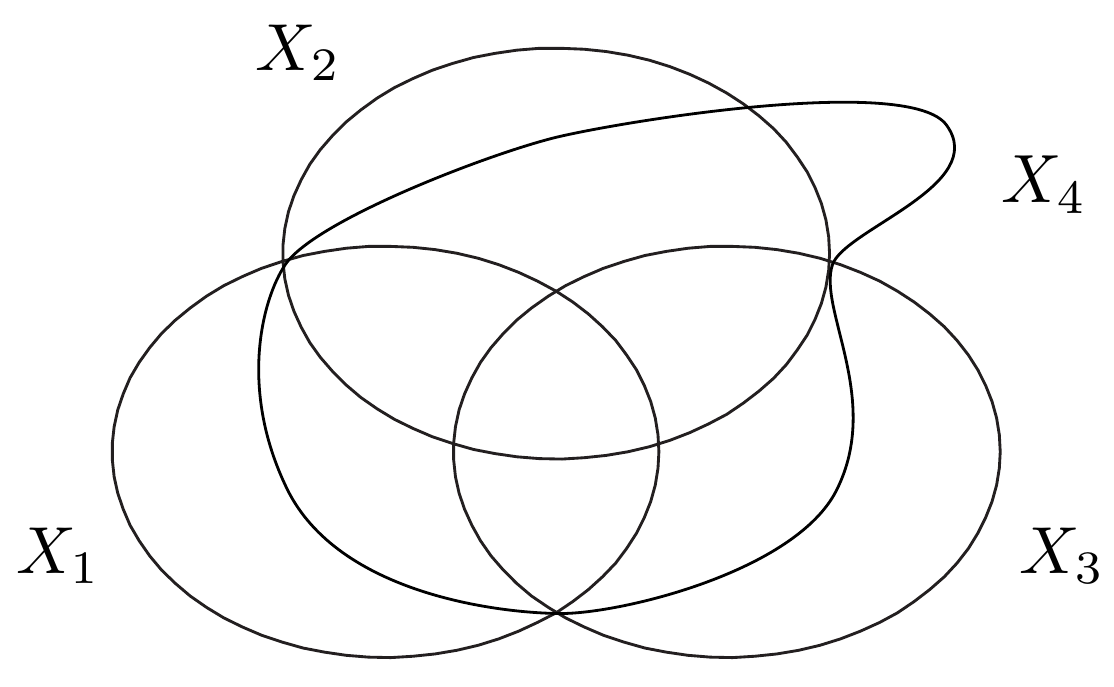}
\caption{The information diagram in Example~\r{eg18}.}
\l{figeg18}
\end{figure}

\newpage

\begin{figure}[h]
\vspace{2.5in}
\centering
\includegraphics[height=2.5in]{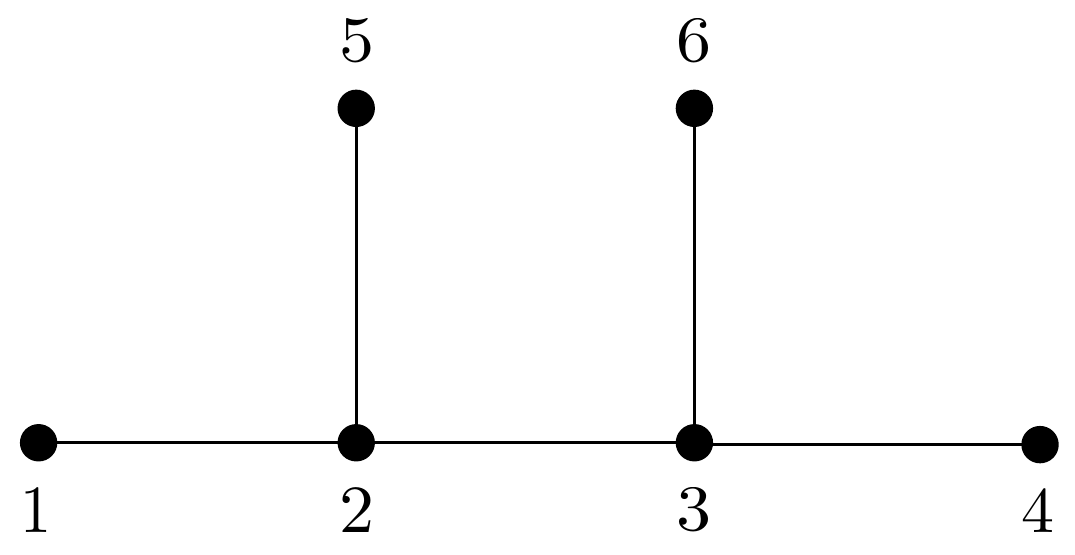}
\caption{The graph $G$ in Example~\r{eg19}.}
\l{figeg19}
\end{figure}

\newpage

\begin{figure}[h]
\vspace{2.5in}
\centering
\includegraphics[height=3in]{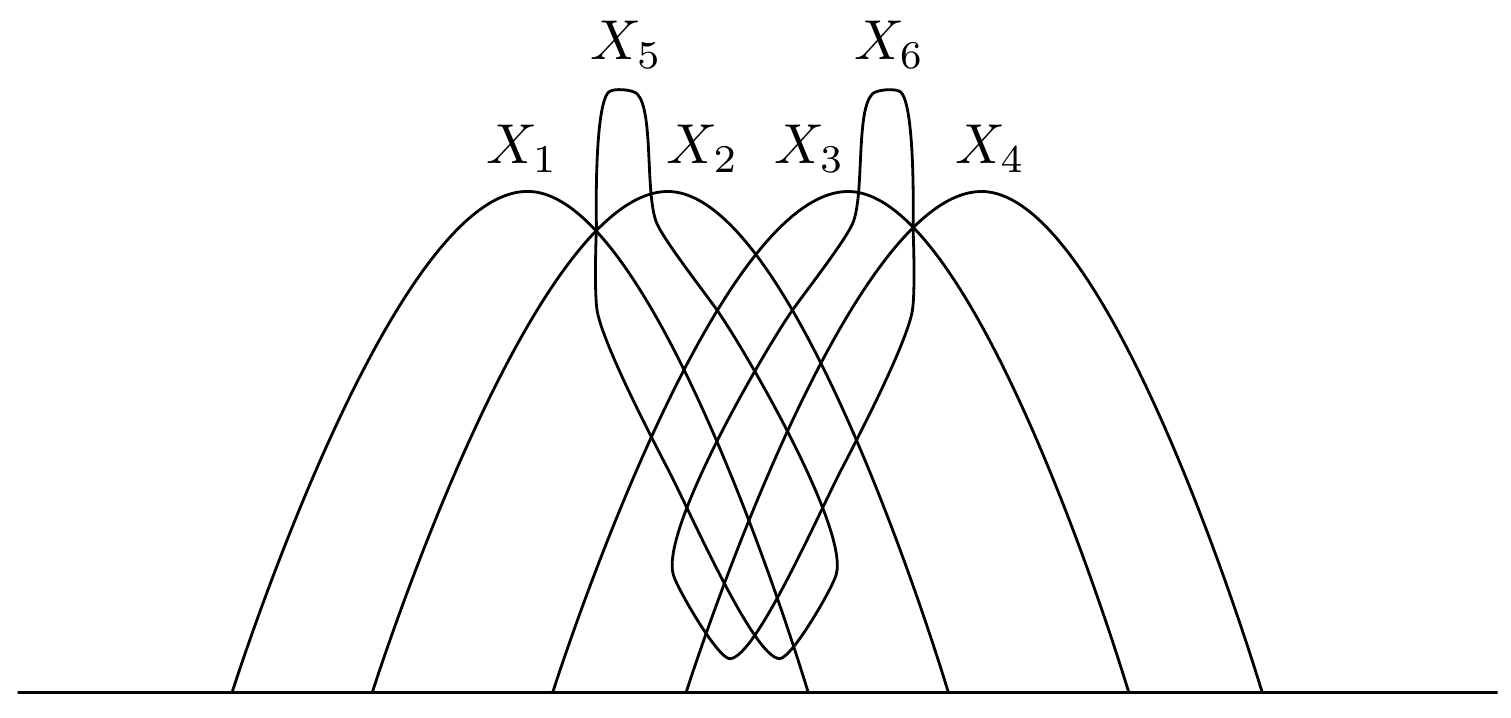}
\caption{The information diagram in Example~\r{eg19}.}
\l{figeg19b}
\end{figure}

\end{document}